\documentclass[prl,aps,amssymb,superscriptaddress,twocolumn, hypertext, showpacs]{revtex4-2} 

\usepackage{hyperref}
\hypersetup{
    colorlinks,
    linkcolor={black},
    citecolor={blue!80!black},
    urlcolor={blue!80!black}
}

\usepackage{graphicx}
\usepackage{dcolumn}
\usepackage{bm}
\usepackage{soul}
\usepackage[utf8]{inputenc}
\usepackage{amssymb, amsthm, amsmath, amsfonts}
\usepackage{upgreek}
\usepackage{multirow}
\usepackage{xcolor}
\usepackage[per-mode=symbol]{siunitx}
\usepackage{braket}
\usepackage{textcomp}
\usepackage{booktabs}
\usepackage{gensymb}
\usepackage[T1]{fontenc}
\usepackage{bbm}
\usepackage{siunitx}
\usepackage{bigints}
\usepackage{amssymb, amsthm, amsmath, amsfonts}

\DeclareSIUnit\gauss{G}
\graphicspath{{/}}
\renewcommand{\vec}[1]{\bf #1}
\renewcommand{\i}{{\rm i}}

\makeatletter
\renewcommand*{\@fnsymbol}[1]{\ensuremath{\ifcase#1\or \dagger \or *\or \ddagger\or
   \mathsection\or \mathparagraph\or \|\or **\or \dagger\dagger
   \or \ddagger\ddagger \else\@ctrerr\fi}}
\makeatother

\begin{document}


\title{Coherent microwave, optical, and mechanical quantum control of spin qubits in diamond}

\author{Laura Orphal-Kobin}
\affiliation{\vspace{0.5em}Department of Physics \& IRIS Adlershof, Humboldt-Universit\"at zu Berlin, Newtonstr. 15, 12489 Berlin, Germany}
\author{Cem Güney Torun}
\affiliation{\vspace{0.5em}Department of Physics \& IRIS Adlershof, Humboldt-Universit\"at zu Berlin, Newtonstr. 15, 12489 Berlin, Germany}
\author{Julian M. Bopp}
\affiliation{\vspace{0.5em}Department of Physics \& IRIS Adlershof, Humboldt-Universit\"at zu Berlin, Newtonstr. 15, 12489 Berlin, Germany}
\affiliation{\vspace{0.5em}Ferdinand-Braun-Institut, Gustav-Kirchhoff-Str. 4, 12489 Berlin, Germany}
\author{Gregor Pieplow}
\affiliation{\vspace{0.5em}Department of Physics \& IRIS Adlershof, Humboldt-Universit\"at zu Berlin, Newtonstr. 15, 12489 Berlin, Germany}
\author{Tim Schröder}
\email{tim.schroeder@physik.hu-berlin.de}
\affiliation{\vspace{0.5em}Department of Physics \& IRIS Adlershof, Humboldt-Universit\"at zu Berlin, Newtonstr. 15, 12489 Berlin, Germany}
\affiliation{\vspace{0.5em}Ferdinand-Braun-Institut, Gustav-Kirchhoff-Str. 4, 12489 Berlin, Germany}

\begin{abstract}
\noindent Diamond has emerged as a highly promising platform for quantum network applications. Color centers in diamond fulfill the fundamental requirements for quantum nodes: they constitute optically accessible quantum systems with long-lived spin qubits. Furthermore, they provide access to a quantum register of electronic and nuclear spin qubits and they mediate entanglement between spins and photons. 
All these operations require coherent control of the color center's spin state. This review provides a comprehensive overview of the state-of-the-art, challenges, and prospects of such schemes, including, high fidelity initialization, coherent manipulation, and readout of spin states. Established microwave and optical control techniques are reviewed, and moreover, emerging methods such as cavity-mediated spin-photon interactions and mechanical control based on spin-phonon interactions are summarized. For different types of color centers, namely, nitrogen-vacancy and group-IV color centers, distinct challenges persist that are subject of ongoing research. Beyond fundamental coherent spin qubit control techniques, advanced demonstrations in quantum network applications are outlined, for example, the integration of individual color centers for accessing (nuclear) multi-qubit registers. Finally, we describe the role of diamond spin qubits in the realization of future quantum information applications.
\end{abstract}

\maketitle


\setcounter{secnumdepth}{2} 
\section{Introduction} 

Quantum information processing applications are expected to have a big impact on industry and society. For example, long distance quantum networks~\cite{KimbleNature2008} would enable secure quantum communication,\cite{basso_basset_quantum_2021,ekert_ultimate_2014} distributed quantum computing,\cite{atature_material_2018,wehner_quantum_2018,nickerson_freely_2014} and enhanced quantum sensing.\cite{degenRevModPhys2017} A quantum node in such a network requires an optically accessible, sufficiently coherent quantum system composed of qubits that can be controlled with high fidelity. Wide-band gap solid-state materials stand out, since they can host a large range of color centers capable of hosting long-lived, optically active spin qubits,\cite{munns_quantum_2023, wolfowicz_quantum_2021} and moreover, provide compatibility with photonic integration and on-chip scalability.\cite{wan_large-scale_2020,SipahigilScience2016,golter_selective_2023,lenzini_diamond_2018,palm_modular_2023,li_heterogeneous_2023,SchroederJOSAB2016,MouradianPRX2015}

Color centers in diamond have been investigated extensively regarding quantum applications since the 1990s, when the nitrogen-vacancy color center's (NV) ground state spin-triplet structure and spin coherence have been optically detected in landmark experiments.\cite{vanOortJoPC1988,GruberScience1997} 
While the NV became initially attractive as an excellent quantum nanoscale sensor, both the NV and group IV vacancy color centers (G4Vs) continue developing as promising resources for quantum information processing.\cite{BorregaardPRX2019,ruf_quantum_2021,chen_building_2020} 
The NV is well-known for its long spin coherence times of more than one second,\cite{abobeih_one-second_2018} but limited in optical coherence due to its susceptibility to local electric field fluctuations causing spectral diffusion. However, recently, spectrally stable photon emission in nanostructures has been demonstrated under resonant excitation.\cite{orphal-kobin_optically_2023} G4Vs exhibit due to their inversion symmetry high optical coherence. Extensive research from 2011 on,\cite{BradacNatComms2019,NeuNJP2011} exploring different types of G4Vs and experimental methods, also lead to an increase of spin-coherence times from sub-microseconds up to the order of tens of milliseconds.\cite{SenkallaArxiv2023,SukachevPRL2017}
Moreover, color centers in diamond provide access to a long lived register of electronic and nuclear spin qubits.\cite{BradleyPRX2019,van_de_stolpe_mapping_2023,parker_diamond_2023} 

 \begin{figure*}
        \centering
        \includegraphics[width=0.9\textwidth]{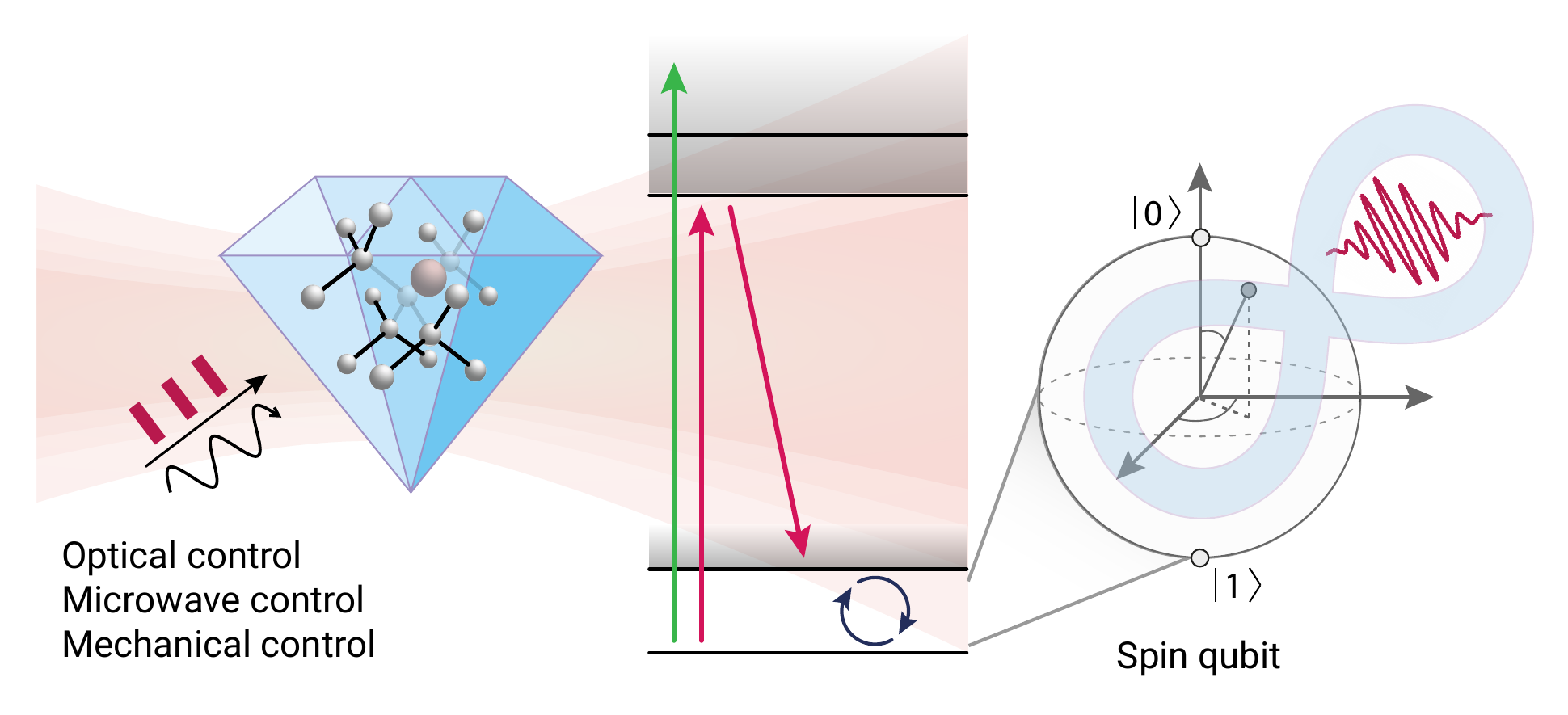}
        \caption{Illustration of the different ingredients of coherent quantum control. Color centers in diamond constitute spin qubits. Initialization, manipulation, and readout of the spin states can be achieved by coherent microwave, optical, and mechanical control methods. Quantum information can be transmitted in flying qubits after spin-photon entanglement generation, e.g., to a distant quantum node in a quantum network.}\label{fig:Fig0}
\end{figure*}

This review is centered on the stack of control operations that are required for any quantum node.\cite{divincenzo_physical_2000} This stack is composed of qubit initialization, its coherent control and readout as well as its ability to mediate entanglement between spins and photons. So far, continuous development enabled spin qubit initialization fidelities of more than $99\%$,\cite{PompiliScience2021} and averaged spin qubit state readout fidelities of more than $97\%$.\cite{HumphreysNature2018} Moreover, color centers in diamond have been used to demonstrate, employing advanced control protocols, spin-photon~\cite{ToganNature2010} and spin-spin entanglement,\cite{BernienNature2013,HumphreysNature2018,knaut_entanglement_2023} the implementation of a three-node quantum network~\cite{PompiliScience2021} as well as qubit teleportation.\cite{hermans_qubit_2022} For entanglement generation, time-bin encoding, to date, has turned out to be the most efficient, however, the NV also allows for polarization encoding.\cite{ToganNature2010} Despite these milestone achievements, there are still persisting challenges that need to be tackled for the implementation of large-scale quantum networks based on color centers in diamond, which will be discussed in the scope of this work as well.

In this review, we aim to provide a concise overview of the current and emerging methods for (coherent) control of spin qubits in diamond towards quantum information processing applications, as indicated in the illustration in \textbf{Figure~\ref{fig:Fig0}}. We do not discuss the role of color centers in diamond for sensing and secure communication. Moreover, for information on other pre-requisites for large scale quantum network applications, such as diamond nanofabrication,\cite{SchroederJOSAB2016} hybrid photonic architectures,\cite{elshaari_hybrid_2020,kim_hybrid_2020,zhu_integrated_2021,riedel_efficient_2023} and frequency conversion to the telecom band~\cite{dreau_quantum_2018,bersin_telecom_2023} we refer to more specialized works and reviews. 

This review is organized as follows: In Section~\ref{sec:2} some fundamental properties of \textit{color centers in diamond} are introduced. The following sections then provide an overview of coherent control techniques. We begin with \textit{microwave control} in Section~\ref{sec:MW} and \textit{optical control} in Section~\ref{sec:opt_ctrl}, which are the most established and are a foundation for most advanced control techniques. We then review more recent methods for realizing single- and two-qubit gates. In particular, we present spin qubit-photon gates mediated by \textit{enhanced light-matter interaction in nanocavities} (Sections~\ref{sec:light-matter} and \ref{sec:optctrl_cavity_spin_manipulation}). In Section~\ref{sec: MechCtrl} we provide an overview of single qubit gates enabled by mechanical spin control, which is mediated by application of dynamical strain. Beyond fundamental control principles, we briefly outline advanced demonstrations in quantum network applications as well as potential future directions of quantum information processing and quantum computing based on color centers in diamond in Section~\ref{sec:perspectives}.

\section{Color centers in diamond}\label{sec:2}

In diamond, more than 500 optically active defects exist~\cite{zaitsev_optical_2001}. For quantum applications color centers with single photon emission and optically accessible spin ground states are most relevant. Here, we focus on the nitrogen-vacancy color center (NV) 
and the negatively charged group IV vacancy color centers (G4V).
There are many more defect centers that have been investigated. Their application range, however, to date has been limited and for a detailed introduction we point at the respective references. Examples are color centers based on magnesium,\cite{corte_magnesium-vacancy_2023} chromium,\cite{AharonovichPRA2010} xenon,\cite{DeshkoLTP2010} and the nickel-nitrogen complex NE8,\cite{GaebelNJP2004} as well as the nickel-vacancy center with near infrared optical activity.\cite{SiyushevNJP2009, WuOptExp2006}

\subsection{Material considerations}

Diamond is a wide-bandgap semicondutor with a bandgap of about 5.5\,eV. Its optical transparency and capability to host a large variety of color centers make it an appealing platform for quantum optics. The color centers discussed in this review are formed from substitutional atoms (impurities) in combination with single lattice vacancies. Within the bandgap, they give rise to a localized atom-like system that can, depending on its charge state, host a long lived spin-qubit, which is amenable to microwave, optical, and mechanical manipulation, and, moreover, optical initialization and readout. In diamond, long spin coherence times can be maintained due to low phonon interaction (high Debye temperature) and low magnetic noise from nearby spins, since it is primarily composed of spin-0 $^{12}$C atoms (around 99$\%$).\cite{zaitsev_optical_2001,markham_cvd_2011}

For quantum operations based on interference of indistinguishable photons, photons emitted into the spectrally narrow zero-phonon line (ZPL) are used. 
The Debye-Waller factor is defined by the ratio of photons emitted into the ZPL, compared to the total emission including the ZPL as well as the spectrally broad and red-detuned phonon-sideband. It varies for different color centers from 3$\%$ for the NV up to 60-80$\%$ for the centrosymmetric G4Vs.\cite{NeuNJP2011, GoerlitzNJP2020} 
The high refractive index of diamond $n=2.41$ leads to a critical angle of $24.5^\circ$ of the total internal reflection at the diamond-air interface, confining most of the emitted photons in the material. 
Fabrication of efficient diamond nanostructures enable high photon collection efficiencies, for example, by collecting photons through an objective (free-space) or by directly coupling to an integrated fiber.\cite{SchroederJOSAB2016, WanNanoLett2018,BurekPRApp2017,torun_optimized_2021} Nanostructures can also be used to increase the spin-photon coupling, which leads to the enhancement of emission into a well defined optical mode (Purcell enhancement, see also Section~\ref{sec:light-matter}), which also improves collection efficiencies. Finally, as a solid-state material, diamond facilitates photonic integration, an essential aspect for scalability, which is a critical factor in the development of quantum technology applications.\cite{lenzini_diamond_2018}

\subsection{Color center formation}\label{sec:formation}

Typically, artificial diamond is used for quantum applications, due to high purity and because it allows for the control of various growth parameters, such as the density of a desired defect species. Diamond can be produced through two primary methods: chemical vapor deposition (CVD) or the high-pressure high-temperature (HPHT) technique. The CVD setup is rather simple and the method allows for flexible control over impurity components and concentrations.\cite{butler_recent_2009} Electronic grade diamond crystals are formed by isotopically purified $^{12}$C, which is obtained from a plasma composed of methane and hydrogen. The $^{12}$C are then deposited onto a substrate, which seeds the growth of the diamond crystal. Defects are introduced by adding a certain concentration of e.g., nitrogen or silicon, into the growth chamber. The addition of impurity components for a well defined time, can even create "delta-doped" layers with precise control of defect center depth at the nanometer scale.\cite{OhnoAPL2012,hughes_two-dimensional_2023} To enhance the formation yield of atom-vacancy color centers, the diamond is exposed to electron irradiation,\cite{AcostaPRB2009} creating lattice damage. While smaller atoms can be incorporated during growth and still maintain good diamond crystal properties, for larger atoms, such as tin and lead, ion implantation into pure electronic grade diamond substrates with subsequent high-temperature (and possibly high-pressure) annealing is applied.\cite{GoerlitzNJP2020,wang_low-temperature_2021,narita_multiple_2023} The annealing causes an atom migration within the diamond lattice, leading to the formation of atom-vacancy complexes and a recovery of the diamond lattice. 

In comparison to defects generated through ion implantation, native (in-grown) single defect centers typically exhibit superior optical properties due to minimal damage to the local lattice environment.\cite{vanDamPRB2019} For large atoms, shallow ion implantation with subsequent overgrowth is a promising approach.\cite{RugarACSNanoLett2020}

Optimising defect formation is still a subject of active research. Co-implantation of atoms and Coulomb-driven defect engineering can lead to improved properties regarding charge-state, spin and optical coherence, as well as increased color center formation yield.\cite{LuehmannNatComms2019,yamamoto_strongly_2013,yurgens_spectrally_2022}
A review on color center generation in diamond can be found in ref.~\cite{smith_colour_2019}.

\subsection{Nitrogen-vacancy color center}

The NV is the most explored color center in diamond. It first made its strides as a platform for quantum sensing of magnetic fields and temperature~\cite{AcostaPRB2009, GrinoldsNatPhys2013, SchirhaglARevPChem2014, JeskeNJP2016, clevensonnatphysics2015} and later emerged as a candidate for quantum information applications. 

The NV is a point defect, that consists of a nitrogen atom and a vacancy, substituting a nearest-neighbor pair of carbon atoms in the diamond lattice. It exhibits C$_{3\mathrm{v}}$ symmetry along the $\left\langle111\right\rangle$ directions, i.e., it is non-inversion symmetric leading to a permanent electric dipole moment which makes the NV energy levels sensitive to electric fields. In its neutral charge state (NV$^0$) with one unpaired electron, it forms a spin $S=1/2$ quantum system. In the negative charge state (NV$^-$), an extra electron is present resulting together with the unbound vacancy electron in a $S=1$ system. From here on in NV refers to the NV$^-$, which, so far, is the most relevant charge state for quantum applications. 

\begin{figure*}
    \centering
    \includegraphics[width=0.85\textwidth]{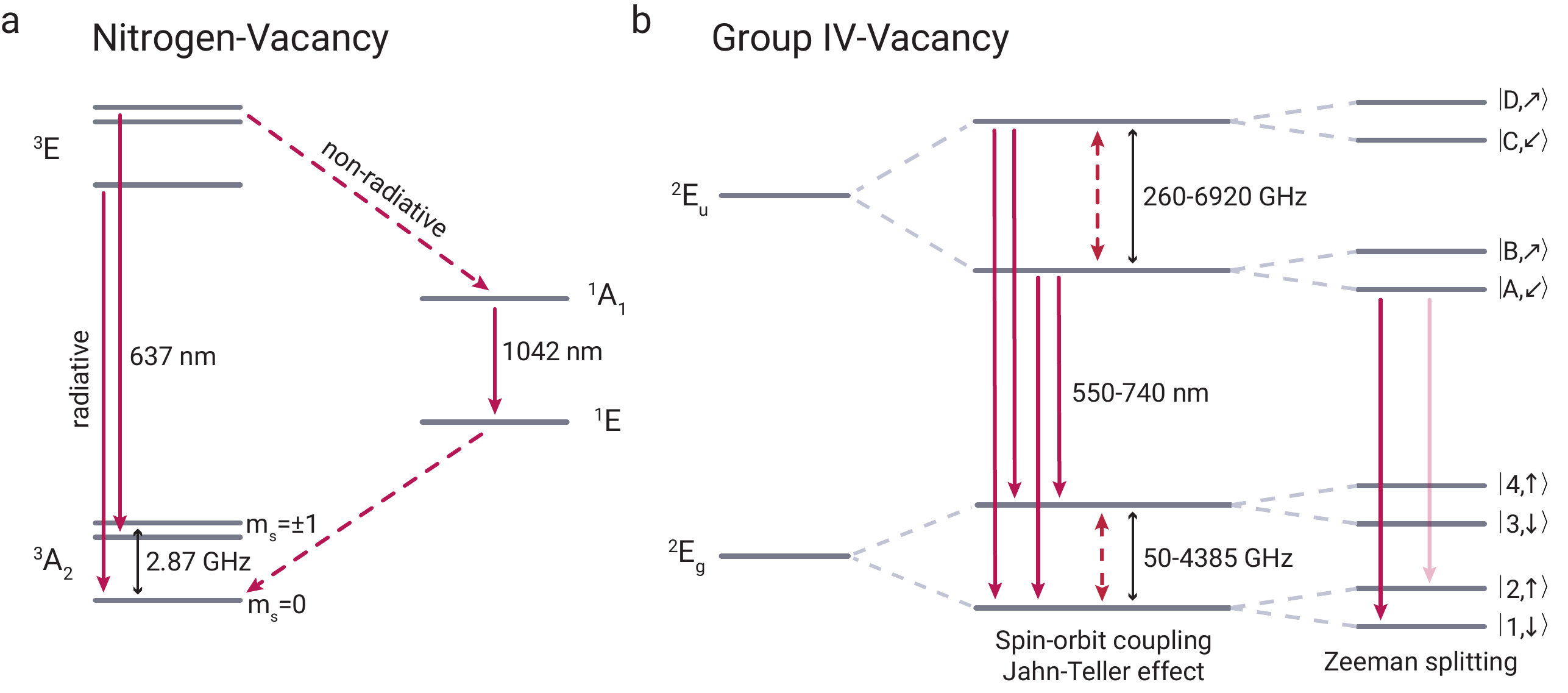}
    \caption{Schematic electronic level structure. (a) Negatively charged nitrogen-vacancy color center ($I = 1$). (b) Negatively charged group-IV vacancy color center ($I = 1/2$).}\label{fig:energylevels}
\end{figure*}

The schematic energy level structure of the NV$^-$ is shown in \textbf{Figure~\ref{fig:energylevels}}(a). Unperturbed, the energy levels of the NV$^-$ involve a ground state ($^3 \textrm{A}_2$), an excited state spin triplet ($^3 \textrm{E}$)~\cite{MansonPRB2006,MazeNJP2011, DohertyPhysRep2013} and two singlets (${}^1\rm A_1$, ${}^1\rm E$).
The ground state spin triplet $^3 \textrm{A}_2$ shows a zero field splitting of $\sim$2.87\,GHz between the $m_{\rm s} = 0$ and $m_{\rm s} = \pm 1$ sublevels. Analogous to the zero field splitting of the ground states the excited state triplet $^3 \textrm{E}$ shows a splitting of $\sim$1.42\,GHz.\cite{batalovPRL2009}
By applying an external magnetic field the energy degeneracy of the $m_s=\pm1$ spin sublevels of the ground state spin triplet can be lifted due to the Zeemann effect. A subset of states of the ground state triplet such as for example $m_{\rm s} = 0$ and $m_{\rm s} = +1$ levels is then typically used as a spin qubit. 

The excited state and ground state manifolds are optically coupled through spin preserving transitions, leading to a total of six ZPL transitions at $\sim637$\,nm. In addition, the two intermediate singlet states ${}^1\rm A_1$ and ${}^1\rm E$ are connected through an optical transition with a wavelength of 1042\,nm.\cite{RogersNJP2008,RogersNJP2015} The two singlet states form another (spin-dependent) decay channel from the excited to the ground states. An electronic decay (intersystem crossing) via these meta-stable states happens predominantly for $m_s=\pm1$ spin-states and are not spin conserving which enables, e.g., spin state initialization through non-resonant optical pumping.\cite{TamaratNJP2008,GoldmanPRL2015} 

NVs formed from naturally the abundant $^{14}$N isotope, which has a nuclear spin $I = 1$ (natural abundance of 99.63$\%$), ion implantation typically relies on $^{15}$N isotopes with nuclear spin $I = 1/2$. Both isotopes naturally lead to two distinct hyperfine structures.
The Debye-Waller factor represents the ratio of photon emission into the ZPL. For the NV only 3$\%$ of the emitted photons are in the ZPL at 637\,nm  and can be potentially considered to be optically coherent and useful for quantum operations, while the large amount is emitted in the phonon sideband (640-800\,nm).

\subsection{Group-IV color centers}

Although the NV has shown exceptional performance in terms of spin coherence properties (see Section~\ref{sec:MW}), its limitations in terms of optical properties (see Section~\ref{sec:opt_ctrl}), have led to the investigation of another family of color centers, known as the G4Vs. These color centers derive their name from the specific group of atoms in the periodic table from which they are composed. G4Vs are formed by an atom, such as silicon,\cite{goss_twelve-line_1996,WangJoPB2006, NeuNJP2011} germanium,\cite{PalyanovSciRep2015, IwasakiSciRep2015} tin,\cite{IwasakiPRL2017, TchernijACSPhot2017, GoerlitzNJP2020} and lead,\cite{wang_low-temperature_2021,trusheim_lead-related_2019} which is positioned at an interstitial site between two vacancy defects in the crystal lattice.\cite{rogers_electronic_2014,wahl_direct_2020} The D$_{3\mathrm{d}}$ point group inversion symmetric arrangement prevents the formation of a permanent electric dipole and therefore inhibits first-order DC Stark Shifts.\cite{DeSantisPRL2021, AghaeimeibodiPRAppl2021}

Completely unperturbed, the electronic structure of the G4V is composed of an $E_g$ ground state level and an $E_u$ excited state level.\cite{HeppPRL2014} Their degeneracy is lifted by the spin orbit and Jann-Teller interaction, leading to four observable optical transitions~\cite{NeuNJP2013,MuellerNatComms2014} as depicted in Figure~\ref{fig:energylevels}(b). Because G4Vs are effective spin 1/2 systems, the levels are two fold degenerate, and can be split through the Zeemann interaction in the presence of a magnetic field. A suitable spin qubit can be chosen, for example, between the two electronic states of the lower ground state branch. The optical selection rules for the four states in the ground state manifold and the four states in the excited state manifold strongly depend on the direction of the magnetic field, and the strength of the spin-orbit interaction. Under the right conditions, the spectral fine structure of G4Vs results in 16 possible optical transitions.\cite{HeppPRL2014}

For G4Vs, more than 50$\%$  of the emission spectrum consists of coherent photons in the ZPL. The Debye-Waller factor increases inversely proportional with the dopant atom size,\cite{ThieringPRX2018} which reduces distortions of the lattice. A comprehensive review of G4V and their application to quantum technologies can be found in ref.~\cite{BradacNatComms2019}.

Interestingly, the ground and excited state branches are coupled by strain, and therefore to the lattice phonon field. Depending on the temperature, phononic processes \cite{JahnkeNJP2015, harris_coherence_2023} can lead to the decoherence of the two lowest branch spin states due to thermal occupation of the phonoic modes. The temperature for which these processes become the leading decoherence mechanism is determined by the splitting of the ground state branches. Depending on the G4V, phonons are frozen out at a few hundred mK (SiV) or at cryogenic temperatures in the range of a few Kelvin (SnV, PbV). The heavier G4Vs, such as the SnV and PbV are much more desirable, due to their longer coherence times at higher temperatures. 

Also for G4V different dopant isotopes can be implanted, leading to a coupling to different nuclear spins.\cite{parker_diamond_2023,harris_hyperfine_2023}

In Table~\ref{tab:fund_prop}, some fundamental properties of different types of color centers in the negative charge state in diamond are summarized. 

\begin{table*}[htb]
    \caption{Summary of fundamental diamond color center properties in their negative charge state.} \label{tab:fund_prop}
    \centering
    \begin{tabular}{lcccc}
    \toprule
    Color center & ZPL resonance & Debye-Waller factor & Lifetime (natural linewidth) & Ground state splitting\\
    \midrule
    NV~\cite{SantoriPRL2006,DohertyPhysRep2013} &  637\,nm  & $3\%$  &   12\,ns (13\,MHz) & 2.87\,GHz \\
    \midrule
    \midrule
    SiV~\cite{collins_annealing_1994,hepp_electronic_2014,RogersNatComms2014,WangJoPB2006} &  737\,nm  & $\sim70\%$  &  $1.2-1.7$\,ns (110\,MHz) & $\sim 50$\,GHz  \\
    GeV~\cite{IwasakiSciRep2015,SiyushevPRB2017,BhaskarPRL2017} &  602\,nm & $\sim70\%$  & $1.1-6$\,ns (45\,MHz) & $150-170$\,GHz\\
    SnV~\cite{GoerlitzNJP2020,IwasakiPRL2017,TchernijACSPhot2017,RugarPRB2019,TrusheimPRL2020,narita_multiple_2023} &  619\,nm  & $\sim60\%$  & $4-8$\,ns (27\,MHz) & $850$\,GHz \\
    PbV~\cite{wang_low-temperature_2021,wang_transform-limited_2023-1,ThieringPRX2018,trusheim_lead-related_2019}&  $550$\,nm  & $\sim30\%$  & 4.4\,ns (36\,MHz)& $3.9 -5.7$\,THz \\
    \bottomrule
    \end{tabular}
\end{table*}

\section{Microwave control} \label{sec:MW}

Since the first demonstrations of optically detected magnetic resonance (ODMR) with NVs in the 1990s,\cite{vanOortJoPC1988, GruberScience1997} color centers in diamond became a well established platform for (nano-)sensors of magnetic fields and temperature~\cite{degenRevModPhys2017,SchirhaglARevPChem2014}. They have since then also proved themselves as promising candidates for quantum network applications.\cite{PompiliScience2021,hermans_entangling_2023} 
The spin state splitting frequencies (under an applied magnetic field) are determined in ODMR experiments.\cite{levine_principles_2019} 
The interaction potential that couples a microwave field to a color center's electrons is typically described by \cite{doherty_theory_2012}
\begin{equation}
    H_{\rm B} = \frac{\mu_{B}}{\hbar} \sum_{i}(\vec l_i + g_e \vec s_i) \vec B \,,
\end{equation}
where $\mu_B$ the Bohr magneton, $\hbar$ the reduced Planck constant, $g_e$ the electron g-factor, $\vec{l}_i = \vec r_i \times \vec p_i$ is the orbital magnetic moment of the $i$th electron, $\vec p_i$ its momentum, $\vec s_i$ its spin, and $\vec B$ the magnetic field used to drive the center. 
The effective Hamiltonians describing the interaction of a magnetic field with the ground state manifold are slightly distinct for the NV and the G4Vs because they have a different spin. For the NV, the effective Hamiltonian can be written in a fairly compact form:
\begin{equation}
    H_{\rm B}^{\rm NV}  = \frac{\mu_B}{\hbar} \vec S \cdot \hat{g}\cdot \vec{B} \,,
\end{equation}
where $\vec S$ is the total spin-1 operator and $\hat{g} = {\rm diag}\{g_\perp, g_\perp, g_\parallel\}$. $g_\perp, g_\parallel$ are the nonaxial and axial effective g-factor components. 

From \cite{hepp_electronic_2014} the effective Hamiltonian for G4Vs becomes
    \begin{align}
        \begin{aligned}
        H_{\rm B}^{G4V}  
            =& \frac{f\gamma_L}{\hbar} B_z\begin{pmatrix}
        0 & i \\
        -i & 0
        \end{pmatrix} \otimes \begin{pmatrix}
        1 & 0 \\
        0 & 1 \\
        \end{pmatrix} 
       + \frac{\gamma_S}{\hbar} \begin{pmatrix}
        1 & 0 \\
        0 & 1 \\
        \end{pmatrix}\\& \otimes  \begin{pmatrix}
        B_z & B_x - \i B_y \\
        B_x+ \i B_y & -B_z \\
        \end{pmatrix}~,
        \label{eq:ham_magnetic}
        \end{aligned}
    \end{align}    
which is written in the basis of the Hilbert space $\mathcal{H} = \mathcal{E}_g \otimes \mathcal{S}_{\rm 1/2}$, where $\mathcal{E}_g$ is spanned by $|e_{gx}\rangle, |e_{gy}\rangle$ and $\mathcal{S}_{1/2}$ by $|\downarrow\rangle, |\uparrow\rangle$. In Equation~\eqref{eq:ham_magnetic}, $\gamma_L = \gamma_s/2$, where $\gamma_s$ is the gyromagnetic ratio, $f$ is a quenching factor, and $\vec{B} = (B_x, B_y, B_z)$. Both these Hamiltonians are the starting point for modelling the interaction of a magnetic field with the vacancy's spin. 

Coupling to a resonant microwave driving field induces rotations on the Bloch sphere, known as Rabi oscillations. Microwave pulses with well-defined duration, power, and phase -- for example a $\pi$-pulse around a certain axis -- can generate controlled spin state population flipping or rotations in the equator plane of the Bloch sphere. Full quantum control can be achieved by control of either the microwave polarization or phase, resulting in the implementation of a set of universal single-qubit gates. By coherent microwave manipulation of the electronic spin and by leveraging the Zeeman interaction, a complete set of single- and two-qubit gates can be engineered operating on two electronic spins. Such control procedures were used to produce entanglement between two electronic spins with high fidelity.\cite{doldeNatComm2014}

\subsection{Electron spin coherence}

To quantify the coherence times of a qubit system, the electron spin-lattice relaxation time $T_1$ (longitudinal relaxation) and spin coherence time $T_2$ (transverse relaxation) have been investigated using microwave control. The $T_1$ time captures the population decay between the two qubit states and sets the fundamental limit for spin coherence, $T_2\le 2\,T_1$. It is determined by phononic processes, such as thermal relaxation by emission, absorption, or scattering of phonons in the diamond lattice. Moreover, depending on the environment also magnetic and electric charge noise can cause random spin-flips.
The spin coherence time determines, how long phase information can be stored in a spin qubit, the memory time of the quantum system.
The $T_2^\star$ time, representing the overall coherence between two states, corresponds to inhomogeneous dephasing processes and is recorded in a Ramsey experiment (free-induction decay).\cite{levine_principles_2019} When the electron spin undergoes precession in the equator plane after being prepared in an equal superposition state ($\pi/2$-pulse), an additional refocusing $\pi$-pulse can decouple the quantum system from quasi-static magnetic fields, protecting the coherence of the spin state. Such a Hahn echo measurement yields the homogeneous $T_2$ time. For comparison, an overview of spin properties in different solid-state materials can be found in ref.~\cite{wolfowicz_quantum_2021}.

Decoherence of the electron spin is mainly caused by the interaction with neighboring spins in the environment, which is a common mechanism for solid-state systems. In diamond, the spin density is relatively low, however, a residual paramagnetic bath is comprised of $^{13}$C spins, P1 centres (nitrogen impurities), and others. The dephasing can be considered as a random magnetic bath field which is directed along the color center's quantization axis. To suppress the interaction of the electron spin with the surrounding spin bath and to prevent decoherence of quantum states, dynamical decoupling (DD) schemes are used. The spin echo pulse sequence can be extended to a train of periodic $\pi$-pulses seperated by a defined delay time $\tau$, $(\tau - \pi - \tau)^N$, called Carr-Purcell-Meiboom-Gill (CPMG) sequence.\cite{CarrPR1954} While CPMG is a single-axis decoupling protocol, decoupling sequences constructed of rotations over two axes preserve robustness against pulse errors and maintain coherence of arbitrary quantum states $\ket{x}$ and $\ket{y}$.\cite{deLangeScience2010,gullion_new_1990} E.g., the XY4 and XY8 protocols are commonly used in which the phase of the pulses is alternated. The coherence time scales with the number of applied pulses $N$. 

Since G4Vs are spin 1/2 systems, they are not only subject to magnetic field fluctuations -- the predominant dephasing mechanism for the NV. Depending on the environment properties, resonant dipolar coupling with the spin bath (e.g. P1 centers) additionally contributes significantly to decoherence. Such a Markovian coupling cannot be mitigated by DD pulses.\cite{BeckerPRL2018}

\subsection{Noise spectroscopy and coupling to nuclear spin registers}

While for the electron spin coherence, nuclear spins in the environment seem to potentially cause detrimental magnetic fluctuations, they can also be used as a powerful resource. Nuclear spins are significantly less affected by magnetic fields in comparison to electron spins due to the difference in the nuclear and Bohr magneton. By coupling a single electron spin to multiple nuclear spins in a controlled fashion, multi-qubit registers can be formed to store quantum information for long times, and be used for quantum algorithms.\cite{TaminiauNatNano2014} Here, the color center acts as an optically active interface qubit and the nearby spins form weakly interacting ancilla qubits, e.g. $^{13}$C and $^{14}$N nuclei for the NV. In DD spectroscopy, the electron spin operates as a quantum sensor to resolve the nuclear-spin environment.\cite{KolkowitzPRL2012,TaminiauPRL2012,ZhaoNatNano2012} Coherent coupling between spins can be mediated through microwave pulse sequences tuned to the respective hyperfine coupling strengths~\cite{ChildressScience2006,DuttScience2007}. Then DD sequences can be applied in order to decouple the target spins from the remaining spin bath. Despite increased quantum memory times and facilitating quantum operations in advanced quantum protocols,\cite{BradleyPRX2019,PompiliScience2021,hermans_qubit_2022} controllable coupling to nuclear ancilla qubits is used for nuclear-assisted high-fidelity electronic spin readout~\cite{jiang_repetitive_2009-1,holzgrafe_error_2019,HopperMicromachines2018} and leads to enhancement in sensing protocols.\cite{LovchinskyScience2016,RosskopfNPJQI2017} 

Recently, a method based on time-asymmetric Hahn echo sequences has been proposed for noise spectroscopy with reduced experimental times and, in turn, more robustness against drifts and accumulation of pulse imperfections compared to DD schemes.\cite{Farfurnik2020}

\subsection{State-of-the-art microwave control experiments}

The NV is famous for its long spin coherence times. Even at room temperature, spin coherence in isotopically purified diamond of $>$1\,ms for the electron spin~\cite{BalasubramanianNatMat2009} and $>$1\,s for the $^{13}$C nuclear spin~\cite{MaurerScience2012} could be preserved. At cryogenic temperatures of 3.7\,K, electron spin relaxation times $T_1$ of one hour and spin coherence times $T_2 > 1$\,s using tailored DD sequences~\cite{abobeih_one-second_2018} have been demonstrated. For NVs coupled to nanocavities, spin coherence times exceeding 200\,µs have been shown.\cite{LiNatComms2015,TrusheimNanoLett2014} For NVs, microwave $\pi$-pulse fidelities of more than $\geq 99\%$ are reported.\cite{PompiliScience2021}

\begin{figure*}
        \centering
        \includegraphics[width=1.0\textwidth]{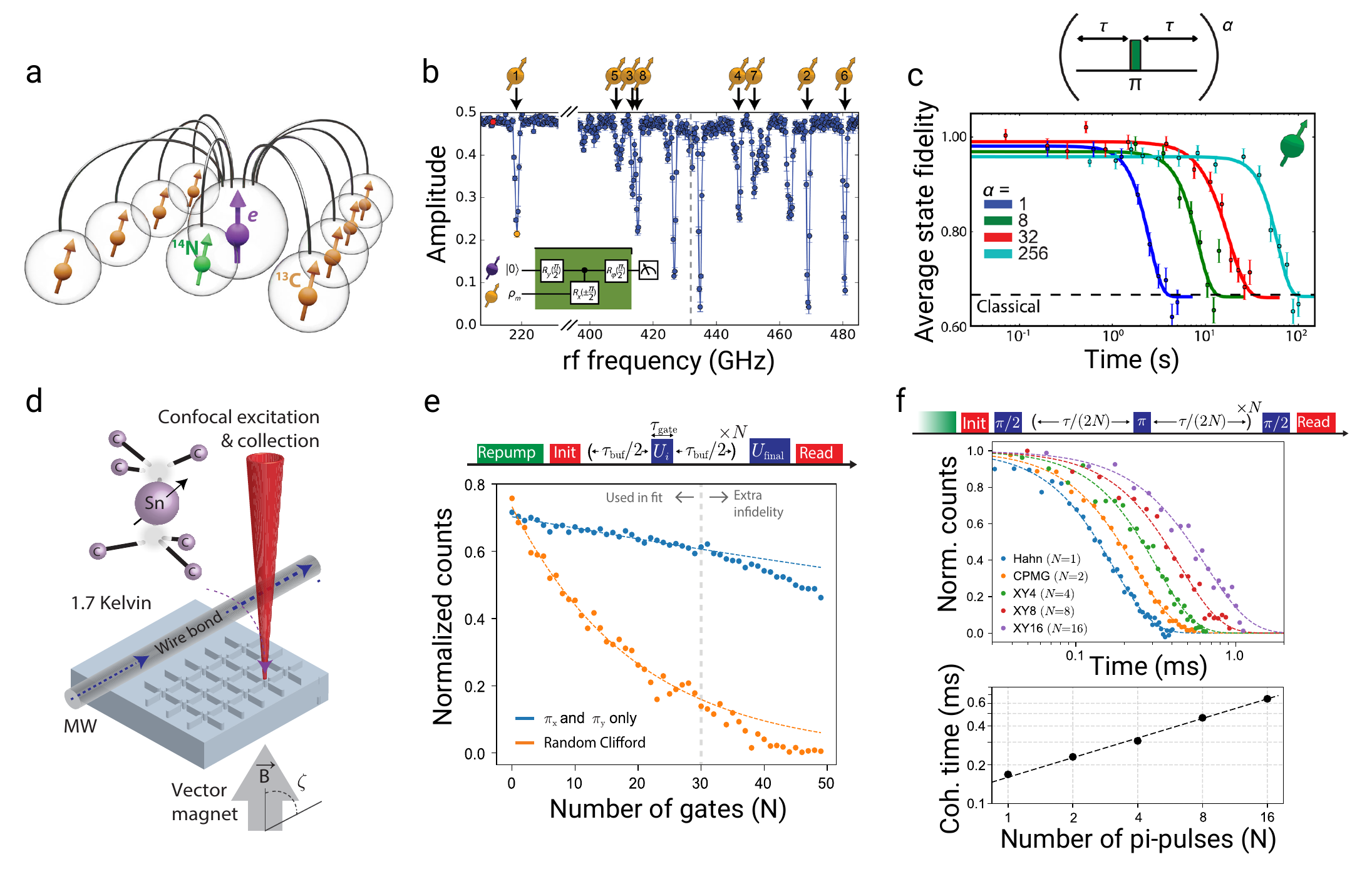}
        \caption{
            Micrwoave control and coherence of spin qubits. Top row: Ten-qubit spin register with an NV as central qubit. (a) Illustration of a multi-qubit register constituted by an NV electron and $^{14}$N nuclear spin coupled to eight $^{13}$C nuclear spins by two-qubit gates. (b) In nuclear spin spectroscopy, the NV electron spin is used as a quantum sensor of its nuclear environment. Here, the electron spin is prepared in a superposition state and a DDrf gate (controlled $\pm\pi/2$-rotation) is applied for different frequencies. If resonant with a nearby nuclear spin, the selective interaction becomes apparent as a dip in the coherence of the electron spin. (c) To confirm that arbitrary quantum states can be protected, the six cardinal states are prepared and the average state fidelity under DD is recorded. Here the decay curves for the $^{14}$N nuclear spin are shown for varying number of DD pulses $\alpha$. With 256 pulses, a state fidelity exceeding the classical memory bound of 2/3 at a time of 75.3\,s is measured.
            Bottom row: Control of an SnV spin qubit. (d) Schematics of the experimental setup using confocal microscopy at 1.7\,K. An SnV spin qubit is incorporated in a diamond nanopillar. Pulsed microwave control is achieved via a wire bond across the sample 60\,µm apart the qubit. The $\braket{100}$~diamond is oriented along the z-axis of the vector magnet. (e) Gate characterization for quantification of spin control fidelities is performed by applying $N$ gates to the qubit, denoted as unitary rotation $U_i$ involving only $\pi$-pulses or a random Clifford gate. Data from the first 30 gates is fitted to the function $\alpha F^N$. A $\pi$-pulse fidelity of $F=99.51\%$ with 54\,ns pulses, and an average random Clifford fidelity of $F=95.04\%$ with 52\,ns pulses have been extracted. (f) The spin coherence time is measured using different DD sequences. Here a $T_2$ time of 650\,µs could be preserved by applying a XY16 protocol (top).  Considering a XY protocol, the coherence time scales with the number of applied decoupling pulses according to $aN^{\chi}$ with $\chi = 0.505$ (bottom). 
            Figures (a)-(c) are rearranged and adapted from Bradley \textit{et al.}~\cite{BradleyPRX2019} under the terms of the Creative Commons Attribution 4.0 International (CC BY 4.0) license. Copyright 2019, American Physical Society. Figures (d)-(f) are rearranged and adapted from Rosenthal \textit{et al.}~\cite{rosenthal_microwave_2023} under the terms of the Creative Commons Attribution 4.0 International (CC BY 4.0) license. Copyright 2023, American Physical Society.
            }\label{fig:MW_control}
    \end{figure*}

Based on magnetic dipolar coupling of NV electron spins, two qubit NV-NV entanglement has been demonstrated at room temperature.\cite{doldeNatPhys2013,doldeNatComm2014} Moreover, in a ten-qubit register consisting of one electron spin and nine nuclear spins, a single-qubit state was preserved for more than 75\,s by using DD sequences on the electron spin and selective phase-controlled driving of the nuclear spins~\cite{BradleyPRX2019} (see \textbf{Figure~\ref{fig:MW_control}}(a)-(c)). Also entanglement between $^{13}$C spin-pairs controlled by an nearby NV has been demonstrated, emphasizing the potential of nuclear spins as qubits.\cite{bartling_entanglement_2022} 

The first demonstrations of microwave control of G4V color center states have been achieved with the `lighter' defect elements, the SiV and GeV color centers, at a few K temperatures.\cite{PingaultNatComms2017,SiyushevPRB2017} In this temperature regime, phononic interactions limit the coherence time. Therefore, further investigations were performed at 100-400\,mK temperatures and moreover DD sequences (CPMG) have been applied to extend $T_2$ times to 13\,ms for an SiV.\cite{SukachevPRL2017} Here, magnetic field noise from the nuclear spin environment was identified as the main decoherence channel. A recent work involving the GeV at 300\,mK benchmarks the state-of-the-art G4V spin control performance using driving $\pi$-pulses with a duration of 77\,ns, which yields T$_{2, \rm CPMG}$ of 24\,ms.\cite{SenkallaArxiv2023} In particular in dilution refrigerators with limited cooling power, drive-induced heat load needs to be considered carefully. When sequences with fast and large amplitude $\pi$-pulses are applied, coherence times may be limited.\cite{BhaskarNature2020}

\begin{table*}[htb]
    \caption{Selection of reported electronic spin qubit coherence times of color centers in diamond in their negative charge state. In the referenced examples, for the NV, the spin qubit is encoded in the $\ket{m_s=0}$ and $\ket{m_s=-1}$ states and for the G4Vs in the lower branch ground states $\ket{1,\downarrow}$ and $\ket{2,\uparrow}$.} \label{tab:MWspin}
    \centering
    \begin{tabular}{lccccc}
    \toprule
    Color center & Temperature & $T_2^\star$ & $T_{2, \rm Hahn}$ & $T_{2, \rm CPMG(N)}$ & $T_{2, \rm XY(N)}$ \\
    \midrule
    \textbf{NV}\cite{muhonen_storing_2014} & 100\,mK   & 160\,µs  & 1\,ms & 0.56\,s (16k)  &  1.58\,s (10k) @ 3.7\,K~\cite{abobeih_one-second_2018} \\
    \midrule
    \midrule
    \textbf{SiV} &    &   &   &   &   \\
    \hspace*{1em}Bulk~\cite{PingaultNatComms2017} & 3.6\,K & 115\,ns & -- & -- & -- \\
    \hspace*{1em}Bulk~\cite{SukachevPRL2017} & 100\,mK & 1.5-13\,µs & -- & 13\,ms (32) & --  \\
    \textbf{GeV}&    &   &   &     &   \\
    \hspace*{1em} Solid immersion lens ~\cite{SenkallaArxiv2023} &  300\,mK  &  1.43\,µs  & 440\,µs  & 24\,ms (8) &  18\,ms (8) \\
    \textbf{SnV} &      &   &   &   &   \\
    \hspace*{1em}Nanopillar~\cite{rosenthal_microwave_2023} & 1.7\,K & 396.6\,ns & 170\,µs & -- & 650\,µs (16) \\
    \hspace*{1em}Membrane~\cite{GuoArxiv2023} & 1.7\,K & 2.5\,µs & 100\,µs & 1.57\,ms (128) & 223\,µs (18) @ 4\,K \\
    \bottomrule
    \end{tabular}
\end{table*}

Coherent microwave control of the SnV spin qubit encoded in the lower ground state branch was also demonstrated (see Figure~\ref{fig:MW_control}(d)-(f)) with $\pi$-pulse fidelities of 99.51\%~\cite{rosenthal_microwave_2023} and 99.36\%.\cite{GuoArxiv2023} As an important pre-requisite for high-fidelity microwave operations for the G4Vs based on heavier group IV defects, i.e. tin and lead, applying strain to the emitters is necessary. Strain introduces orbital mixing and suppresses dephasing processes at temperatures $>4$\,K. The interplay of strain and spin-orbit coupling is a topic of ongoing investigations. Recent theory works identify regimes to suppress phonon-mediated decoherence by changing magnetic-field and strain bias to allow higher temperature operation,\cite{harris_coherence_2023} and there are indications that for certain magnetic field configurations, strain is not only unnecessary but can limit the efficiency of the coherent control.\cite{pieplow_efficient_2023} Finally, the larger dopants are promising candidates for increased coherence times at temperatures in the K regime. For example, the PbV center is predicted to exhibit a spin coherence time on the order of a millisecond at 9\,K.\cite{wang_low-temperature_2021}

While the NV's ($S=1$) resonance condition is in first order sensitive to the particular coupling terms between two spins, G4Vs ($S=1/2$) couple only weakly to nuclear spins. However, initialization and read-out of a $^{13}$C nuclear spin was demonstrated for the first time by using the SiV.\cite{MetschPRL2019,NguyenPRL2019} In addition, nuclear spins of G4V dopant isotopes can be accessed (even at zero-magnetic fields).\cite{parker_diamond_2023,harris_hyperfine_2023}

The reported microwave spin coherence times for the different color centers are listed in Table~\ref{tab:MWspin}. For the NV, the spin qubit encoding is usually implemented in the ground state spin states $\ket{m_s=0}$ and $\ket{m_s=\pm 1}$. For the G4V, in the referenced examples, the lower branch ground states $\ket{1,\downarrow}$ and $\ket{2,\uparrow}$ have been used. 

\section{Optical control} \label{sec:opt_ctrl}

Qubit state initialization via optical pumping and optical single-shot readout are key operations in quantum control protocols, such as spin-photon entanglement generation. In fundamental research, optical control is performed in a confocal microscopy setup. At low temperatures, the spin-selective optical transitions can be spectrally resolved and addressed. While off-resonant excitation or optical pumping resonant to the color centers' neutral charge state can be used for re-initialization in their negative charge state,\cite{aslam_photo-induced_2013,SiyushevPRL2013} here we present resonant excitation and Raman schemes for optical control between ground and excited states for deterministic photon retrieval, and for optical control of the ground state spin.

For initialization and single-shot readout of spin qubits, resonant excitation schemes are applied and the state-dependent fluorescence is recorded. In particular, for entanglement generation the coherent photons emitted into the ZPL are collected. Separation techniques of excitation laser light and single signal photons are cross-polarization, temporal filtering, and spatial filtering.

\subsection{Optical coherence and resonant control schemes}

By the application of laser pulses resonant to the ZPL transitions (orbital transitions), Rabi oscillations are driven through the interaction with the electric field 
\begin{equation}
    H_{\rm El.} =  - \textbf{d}\cdot \textbf{E}\,,
    \label{eq:resonant_rabi}
\end{equation}
leading to the generalized Rabi-frequency 
\begin{equation}
    \Omega ' =  \sqrt{\Omega^2 + \Delta ^2}\,,
\end{equation}
where $\Omega = \textbf{d}\cdot \textbf{E}/\hbar$ is the resonant Rabi frequency, {$\textbf{d}$} corresponds to the transition dipole moment, $\textbf{E}$ to the electric field vector, and $\Delta$ to the detuning from the transition resonance.\cite{fox_quantum_2006} From a Rabi measurement, the corresponding $\pi$-pulse parameters, i.e. excitation power and pulse duration, can be directly extracted. The pulse area is defined by $\Theta = \int \Omega (t)\,$d$t$.

In general, the optical coherence is limited by spontaneous emission, phonon-induced relaxation processes, spin-mixing and spectral diffusion. Depending on the color center and type of sample, the impact of these processes will be different. The decoherence mechanisms prevent high initialization and qubit gate fidelities. 

    \begin{figure*} 
        \centering 
        \includegraphics[width=.8\textwidth]{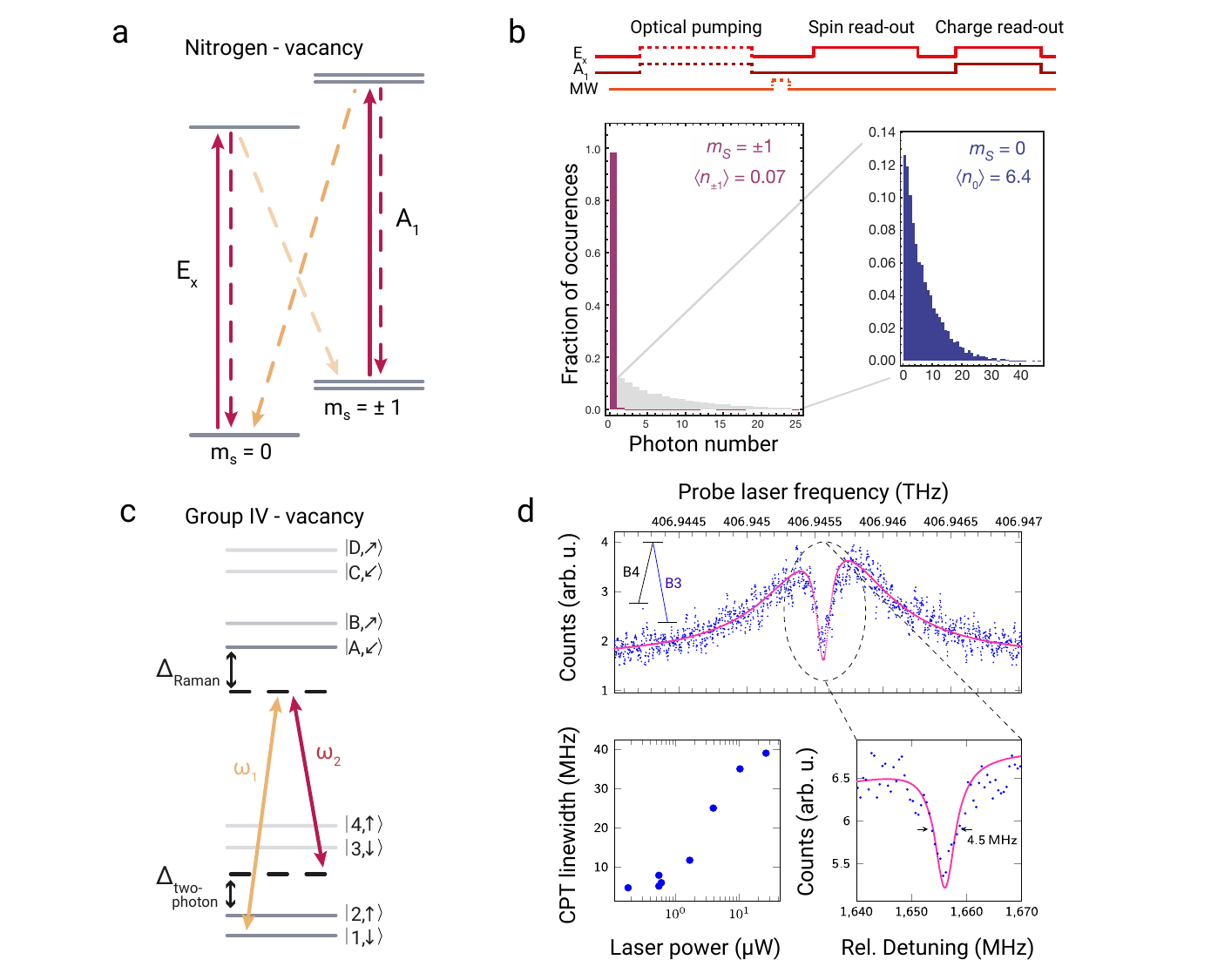}
        \caption{
            Optical control. (a) Simplified level scheme of the NV color center showing the optical $E_x$ and $A_1$ transitions. Single shot readout is performed by resonant excitation of the $E_x$ transition (left solid red line) which results with large probability in photon emission (left dashed red line). With low probability, spin-mixing occurs (light yellow dashed line). For charge state initialization in the ground ${m_s=0}$ spin state, the $A_1$ transition is optically pumped. Non-spin conserving transitions via the metastable singlet states (infared, not shown here) enable spin state initialization with high fidelities ${>99.7\%}$. (b) top: Pulse sequence for electronic spin readout. After charge state initializtaion (not shown) either the $A_1$ or $E_x$ transition is optically pumped for initialization in the ${m_s=0}$ or ${m_s=\pm1}$ spin state. After optional microwace manipulation, the spin state is readout by a laser pulse resonant to the $E_x$ transition. bottom: Photon count statistics integrated over 10.000 measurements during an electronic spin readout pulse with 100\,µs duration after initialization in the ${m_s=\pm1}$ (red) or ${m_s=0}$ (grey, inset) spin state. (c) Optical quantum control of the spin levels through a lambda configuration. When ${\Delta_{\rm two-photon}=0}$, i.e. two-photon resonance is achieved, a coherent population transfer via the excited state without ever occupying it (Raman scheme) can be carried out. When ${\Delta_{\rm Raman}=0}$, the coherent population trapping (CPT) effect occurs involving the ground state (spin) levels and inhibits excitation processes. The linewidth of the two-photon resonance as well as the efficacy of the population trapping (fluorescence contrast) can be analyzed to acquire the rates of the incoherent processes that affect the driven levels. (d) An example CPT measurement with a SiV color center. One of the lasers is stabilized on the B4 transition and a second laser frequency scans around the B3 transition. When both lasers are in resonance, the fluorescence significantly reduces. A characteristic CPT dip is obtained from this measurement. By varying the laser powers, it is possible to estimate the dip's linewidth at zero power, which corresponds to the dephasing rate of the investigated levels. The transition labels in the inset are adapted to the level scheme in (c).
            Figure (b) is adapted and reprinted from Robledo \textit{et al.}~\cite{RobledoNature2011} with permission. Copyright 2011, Nature Publishing Group. Figure (d) is reprinted from Rogers \textit{et al.}~\cite{RogersPRL2014} under the terms of the Creative Commons Attribution 3.0 International (CC BY 3.0) license. Copyright 2014, American Physical Society.
        }\label{fig:optctrl}
    \end{figure*}

For the NV's $E_x$ transition connecting the ground and excited $m_s=0$ states, more than 10 oscillation cycles have been resolved at an excitation power of 38\,µW with Rabi frequencies of $\Omega = 2\pi \times 410$\,MHz corresponding to a $\pi$-pulse duration of 1.2\,ns duration.\cite{robledo_control_2010} In early works, paving the way for the first remote spin-spin entanglement demonstration based on NVs,\cite{BernienNature2013} high fidelity spin qubit state preparation, manipulation, and readout have been demonstrated.\cite{RobledoNature2011} State initialization in the $m_s=0$ ($m_s=\pm 1$) spin ground state can be achieved by resonant optical pumping of the $A_1$ ($E_x$) transition between the ground and excited $m_s=\pm1$ ($m_s=0$) spin states with a high (low) probability of relaxation via the non-spin conserving metastable infrared singlet state (spin-mixing in the excited states), as shown in \textbf{Figure~\ref{fig:optctrl}}(a). Here, a preparation error of the $m_s = 0$ ground state of $(0.3 \pm 0.1)\%$ was extracted. The performance of resonant single-shot readout is characterized after spin state initialization by resonant excitation of the $E_x$ transition. While for initialization in the $m_s=\pm 1$ spin state, negligible numbers of detected photons are expected, for the initial $m_s=0$ state on average six photons were detected during the readout window with optimal duration of 40\,µs, resulting in an average readout fidelity of $F_{\rm avg}=(F_{m_{\rm s}=0}+F_{m_{\rm s}=\pm1})/2=(93.2\pm 0.5) \%$ (see example photon count histograms with 100\,µs readout window in Figure~\ref{fig:optctrl}(b)). For entanglement generation, usually optical $\pi$-pulses resonant to the $E_x$ transition with a duration of about 2\,ns are used. In advanced protocols after delivery of entanglement, the NV state is readout in a particular basis with fidelities of $>95\%$ for the bright $m_{\rm s}=0$ ground state and with $>99.5\%$ for the $m_{\rm s}=\pm 1$ dark state employing methods as described above.\cite{HumphreysNature2018} 
In a low-strain NV environment, spin-mixing and phonon-induced transitions within the excited states while driving the $E_x$ transition are significantly reduced. A completely different approach for spin-state readout is based on photoelectric readout by spin-to-charge conversion.\cite{zhang_high-fidelity_2021,siyushev_photoelectrical_2019} This method allows for suppression of spin-flip errors and support fully integrated quantum devices. Single-shot averaged readout fidelities $>95\%$ with a 10\,µs spin-to-charge conversion duration have been demonstrated for an NV electron spin in the presence of high strain and fast spin-flip process.\cite{zhang_high-fidelity_2021} 

Also for G4V color centers, optical spin state initialization and readout have been demonstrated. To access the electronic spin optically, an external magnetic field with slight misalignment with respect to the color center's symmetry axis is applied, inducing spin state mixing.\cite{PingaultPRL2014} 
At low temperatures on the order of hundreds of mK, the lower orbital branch of the ground state is effectively decoupled from the phonon bath, i.e. thermalization is inhibited, resulting in an orbital polarization in the lower branch. Initialization can be performed by spin pumping: by selective excitation of one of the spin states targeting any of the excited states will populate the other spin state due to radiative decay from the excited state to the other spin state when a misaligned magnetic field is applied. If the highest lying excited state branch is targeted, fast phononic decay processes result in a similar decay chain and therefore also allow initialization. 
When the spin is fully initialized, the excitation field cannot drive transition cycles anymore and the color center is in a dark state.
It has been demonstrated that, e.g. by driving the D1 transition, a G4V color center can be initialized into the spin-up ground state $\ket{2,\uparrow}$~\cite{PingaultNatComms2017} (see level scheme in Figure~\ref{fig:energylevels}(c)). 

The initialization fidelity can be extracted by recording the time-resolved fluorescence during the excitation
pulse. The spin polarization is determined by the ratio of the optical pumping rate to the spin relaxation rate.\cite{SiyushevPRB2017} By applying a second laser pulse resonant to the first initialization pulse and by measuring the recovery of the fluorescence, the spin can be read out. A comparison between the fluorescence peak areas of the initialization and readout pulses yields the spin polarization.\cite{PingaultNatComms2017} 
High fidelity spin state single-shot readout can be achieved when high, spin-flip free, cyclicity ($\propto \langle i| {\bf d}\cdot {\bf E}| j \rangle$ with $| i\rangle$, $| j \rangle$ being the involved levels) of the spin conserving transition is maintained, e.g. driving transition $\sim 10^5$ times. The required cyclicity is obtained by aligning the angle of the applied magnetic field with the G4Vs symmetry axis.\cite{SukachevPRL2017} However, even for large magnetic field misalignment of $54.7\degree$ between the magnetic field and the symmetry axis of the defect, a reasonable cyclicity of spin-conserving transitions could be achieved, allowing for a readout fidelity of $74\%$, which surpasses the threshold of $50\%$ for meaningful readout implementations.\cite{Gorlitznpj2022} Here, the spin was initialized via the spin-conserving $B2$ transition and for readout a laser pulse resonant to the $A1$ transition was applied.

Rabi oscillations and $\pi$-pulse excitation with G4Vs have been demonstrated.\cite{BeckerNatComms2016, arjona_martinez_photonic_2022} At large driving powers $P/P_{\rm sat}=364$, an optical $\pi$-rotation was performed by a $\pi$-pulse of 1.71\,ns duration (much faster than $T_1$) with $77.1\%$ fidelity.\cite{arjona_martinez_photonic_2022} However, the spin state degeneracy was not lifted by applying a magnetic field.

In Table~\ref{tab:G4V_opt} selected reported spin-state initialization and readout fidelities for the SiV, GeV, and SnV are summarized.

In addition to the presented resonant methods, a novel coherent excitation scheme, called Swing-UP of the quantum EmitteR population (SUPER), \cite{Bracht2021} has been developed. By using high intensity, two-color, ultrashort, and non-resonant pulses, it is possible to rotate the qubit vector around two different Bloch sphere axes simultaneously such that their combined effect results in a full inversion. This scheme has been demonstrated with quantum dots~\cite{KarliNanoLetters2022, BoosArxiv2022} and recently, for the first time with color centers in diamond~\cite{torun_super_2023}. The non-resonant nature of the scheme enables spectral filtering of the excitation field and detection of the coherent single photons in the ZPL.

\begin{table*}[htb]
    \caption{Selection of reported spin-state initialization and readout fidelities for G4Vs in their negative charge state based on optical control methods. The magnetic field angle describes the angle between the applied magnetic field direction and the G4V symmetry axis. The labeling of the involved transitions matches the labeling in Figure~\ref{fig:energylevels} and Figure~\ref{fig:optctrl}. } \label{tab:G4V_opt}
    \centering
    \begin{tabular}{lccccccc}
    \toprule
    \multicolumn{1}{c}{} & \multicolumn{2}{c}{\textbf{SiV}} & \multicolumn{2}{c}{\textbf{GeV}} & \multicolumn{2}{c}{\textbf{SnV}} \\
    \cmidrule(rl){2-3} \cmidrule(rl){4-5} \cmidrule(rl){6-7}
    \textbf{  } & {\cite{PingaultNatComms2017}} & {\cite{SukachevPRL2017}} & {\cite{SiyushevPRB2017}} & {\cite{SenkallaArxiv2023}} & {\cite{Gorlitznpj2022}} & {\cite{GuoArxiv2023}} \\
    \midrule
    \textbf{Settings} &    &   &   &   &   &   \\
    \hspace*{1em}Temperature & 3.6\,K & 100\,mK & 2\,K & 300\,mK & 2\,K & 1.7\,K \\
    \hspace*{1em}Magnetic field (strength, angle) & $109.5\degree$ & 2.7\,kG, $<0.5\degree$ & 0.3\,T, $54.7\degree$ & 0.1\,T, $0\degree$ & 0.2\,T, $54.7\degree$ & 81.5\,mT, $0\degree$ \\
    \textbf{Initialization} &    &   &   &   &   &   \\
    \hspace*{1em}Transition & $D1$ & $A1$, $B2$ & -- & $B2$ & $B2$ & $A1$ \\
    \hspace*{1em}Pulse duration & 500\,ns & 30\,ms & -- & 1\,ms & 200\,µs & 24.2\,µs \\
    \hspace*{1em}Spin polarization & $85\%$ & -- & $59\%$ & $98\%$ & $98.9\%$ & $98.82\%$\\
    \textbf{Readout/gate operation} &    &   &   &   &   &   \\
    \hspace*{1em}Transition & $D1$ & $A1$ & -- & -- & $A1$ & --\\
    \hspace*{1em}Pulse duration, readout window & 500\,ns, 30\,ns & 150\,ms, 20\,ms & -- & -- & 200\,µs & -- \\
    \hspace*{1em}Fidelity (average) & -- & $89\%$ & --& -- & $74\%$ & -- \\
    \bottomrule
    \end{tabular}
\end{table*}

\subsection{Coherent Raman control}

While microwave control methods are effective for high-fidelity quantum operations, they face challenges such as limited scalability,\cite{brecht_multilayer_2016} heating,\cite{wang_microwave_2022} and a limited frequency bandwidth (1 GHz - 300 GHz).\cite{jones_national_2013} This has led to exploring optical control methods as an alternative for coherently controlling quantum states. Since the energy difference in a spin qubit typically doesn't fall within the optical spectrum, optical control must be engineered using so called Raman control techniques, which are sometimes summarized under the process of stimulated Raman adiabatic passage (STIRAP).\cite{vitanov_stimulated_2017} These schemes can be used for systems that have a lambda level structure, where two chosen long-lived quantum states are optically coupled to an energetically higher excited state with a comparatively short lifetime as shown in Figure~\ref{fig:optctrl}(c). In reality, states and couplings outside the lambda level configuration can interfere with the desired Raman control operation. Because any other unwanted optical interaction may introduce errors in the quantum gate operations, they have to be carefully accounted for when designing such a control method.\cite{TakouPRB2021, vezvaee_avoiding_2023} 

So called coherent population trapping (CPT) schemes \cite{Arimondo1976} are a subset of STIRAP. In CPT, two optical transitions within a lambda system are driven simultaneously by two lasers, fulfilling the two-photon resonance condition $\Delta_{\rm two-photon} = 0$. Through a quantum interference effect, one of the adiabatic eigenstates of the driven system no longer contains the short-lived excited state. This allows for the optical control of the spin qubit without any involvement of the short-lived excited state, which would otherwise reduce the fidelity of the desired operation. 

In the rotating wave approximation, the lambda system's Hamiltonian is typically written as (using for example only the states $|1,\downarrow\rangle$, $|2,\uparrow\rangle$, $|A,\downarrow\rangle$ as a basis, (see Figure~\ref{fig:optctrl}(c)) \cite{vitanov_stimulated_2017}: 
\begin{equation}
    H(t)/\hbar = 
    \begin{pmatrix}
    0 & 0 & \Omega_1(t)/2 \\
    0 & \Delta_{\rm two-photon} & \Omega_2(t)/2 \\
    \Omega_1(t)/2 & \Omega_2(t)/2 & \Delta_{\rm Raman}
    \end{pmatrix}  \,,
    \label{eq:RWA_raman}
\end{equation}
where $\Omega_{1,2}$ is given by Equation~\eqref{eq:resonant_rabi}, $d\rightarrow d_{1,2}$ are the respective dipole transition operators (assumed to be real-valued), $\vec{E} \rightarrow \vec{E}_{1,2}(t)$, which are the time-dependent amplitudes of two lasers with central frequencies $\omega_{1,2}$. The two-photon detuning is given by $\Delta_{\rm two-photon} = \omega_{1}-\omega_{2}$ and the single photon detuning is $\Delta_{\rm Raman} = \omega_{\rm A1}-\omega_{1}$ with $\omega_{\rm A1}$ as the optical transition frequency. 
The authors of \cite{TakouPRB2021} used the Hamiltonian in Equation~\eqref{eq:RWA_raman} as a basis for theoretically designing fast Raman gates which act on the ground state qubit of the SiV and SnV. They simulated gates exceeding a fidelity of 99$\%$ for pulses with $20$\,ps duration. The method was further refined in \cite{vezvaee_avoiding_2023} addressing leakage errors. Another theoretical proposal for a double resonant fractional STIRAP was proposed in \cite{lacour_implementation_2006} which employs a complex sequence of pulses required to remove the state-dependent outcome of the single resonant pulse sequence. 

For a large single photon detuning $\Delta_{\rm Raman}$ and $\Delta_{\rm two-photon} = 0$, Equation~\eqref{eq:RWA_raman} can be approximated by adiabatic elimination of the excited state to become \cite{vitanov_stimulated_2017}
\begin{equation}
    H_{\rm eff}(t)/\hbar = 
    \begin{pmatrix}
    0 &  \Omega_{\rm eff}(t)/2 \\
    \Omega_{\rm eff}(t)/2 & \Delta_{\rm eff}(t)
    \end{pmatrix}  \,,
    \label{eq:RWA_raman_eff}
\end{equation}
where 
\begin{align}
    \begin{aligned}
    & \Omega_{\text {eff }}(t) =-\frac{\Omega_1(t) \Omega_2(t)}{ \Delta_{\rm Raman}}\,, \\
    & \Delta_{\text {eff }}(t) = \frac{\Omega_2(t)^2 - \Omega_1(t)^2}{2 \Delta_{\rm Raman}}\,.
    \end{aligned}
\end{align}
The time-dependent effective Hamiltonian $H_{\rm eff}(t)$ generates rotations of the qubit on the Bloch sphere. By extending Equation~\eqref{eq:RWA_raman} to include the relative phase of the two Raman lasers, rotations around the $x$ and $y$-axis of the Bloch sphere can be engineered and therefore arbitrary SU(2) operations.\cite{hamada_minimum_2014}   

CPT may also be employed spectroscopically to investigate the decoherence time of a qubit. For this purpose, one laser is set to match one of the transitions ($\Delta_{\rm Raman}=0$) and then $\Delta_{\rm two-photon}$ is varied by tuning the other laser's frequency while the spontaneous emission (fluorescence) of the system is monitored. When $\Delta_{\rm two-photon} = 0$ (two-photon resonance), a noticeable decrease or `dip' in the fluorescence signal occurs, because the system is driven into the dark state, where the population is coherently trapped in the ground states. The linewidth and contrast of the fluorescence dip depends on the coherence of the probed levels, which can be determined by analyzing the dip's width and depth. 

The initial implementation of CPT for analyzing a diamond spin qubit, specifically an NV color center, was reported in \cite{SantoriPRL2006}. The authors found an effective spin qubit ground state decoherence time of 833\,ns in a temperature range between 2 and 10\,K. In later experiments with G4V color centers, T$_2^*$ dephasing times were found to be 45\,ns at 4\,K \cite{PingaultPRL2014} for the lower branch $\ket{1}$-$\ket{2}$ and 35\,ns at 5\,K \cite{RogersPRL2014} for the upper branch $\ket{3}$-$\ket{4}$ SiV spin levels (Figure \ref{fig:optctrl}(d)), 19\,ns at 2\,K for the lower branch GeV spin levels~\cite{SiyushevPRB2017} and 5\,µs at 1.7\,K for lower branch SnV spin levels.\cite{Gorlitznpj2022} One similar application of dephasing measurements with CPT showed that applying strain can extend coherence times from 40\,ns to 250\,ns for a lower branch SiV spin.\cite{SohnNatComms2018} 
Other applications of this effect include acquiring the CPT dip as a two-photon resonance identification method with SnVs as an initial step for optical control.\cite{DebrouxPRX2021} 

The first experimental demonstration of an optical STIRAP scheme with a color center was performed with an NV \cite{GolterPRL2014} and involved controlling the m$_{\rm s}=\pm1$ levels. This was a remarkable result as the splitting of these levels is usually below 500\,MHz under the typical magnetic fields used in experiments, and are inaccessible by microwave generators. The required two pulses were generated using acousto-optic modulators (AOM). Employing $\sim$\,600\,ns $\pi$-pulses, a T$_2^*$ time of 1\,µs at 5\,K was measured. In addition, adiabatic passage by engineering pulses with varying rise times was also investigated. The fluorescence responses with relative pulse delays were measured and found to be in agreement with simulated values. In \cite{chu_all-optical_2015} another method was developed, which involved a single detuned pulse generated from an electro-optic modulator (EOM). The pulse covered both frequency components due to its broadband nature, enabling faster gates with a duration of 25\,ns and a fidelity of $>$85\% at 7\,K.

The STIRAP method was also applied to G4Vs. For example, the SiV ground state orbital levels were controlled with an ultrashort 1\,ps single broadband pulse \cite{BeckerNatComms2016} with 30\% fidelity at 5\,K. The fidelity was limited by the experimental constraints as the control pulse had to be applied after the SiV had already been partially thermalized. In a later work,\cite{BeckerPRL2018} spin control was demonstrated at 40\,mK by freezing out phonons. Two required frequency components were obtained by sideband generation from an EOM. By cascading with an AOM, 8\,ns $\pi$-pulses were generated and T$_2^*= 29$\,ns, T$_{\rm 2,Hahn}= 138$\,ns were obtained. In the latest G4V optical control work,\cite{DebrouxPRX2021} an SnV was controlled similarly employing EOM sideband generation. A 139\,ns $\pi/2$-pulse with 90.9\% fidelity was generated. Higher rotations were not achievable due to laser power limitations. Decreasing the detuning was not possible either as both the gate fidelities and coherence times were limited by the optical scattering from the excited state. A T$_2^*= 1.3$\,µs, T$_{\rm 2,Hahn}=28.3$\,µs, and T$_{\rm 2,CPMG4}= 300$\,µs were obtained by applying optical refocusing pulses.

An early review of optical control methods for the SiV can be found in ref.~\cite{BeckerPSSa2017}.

\subsection{Spectral diffusion, coherent photons, and spectral control} \label{sec:spectral_control}

\begin{figure*}
        \centering
        \includegraphics[width=0.8\textwidth]{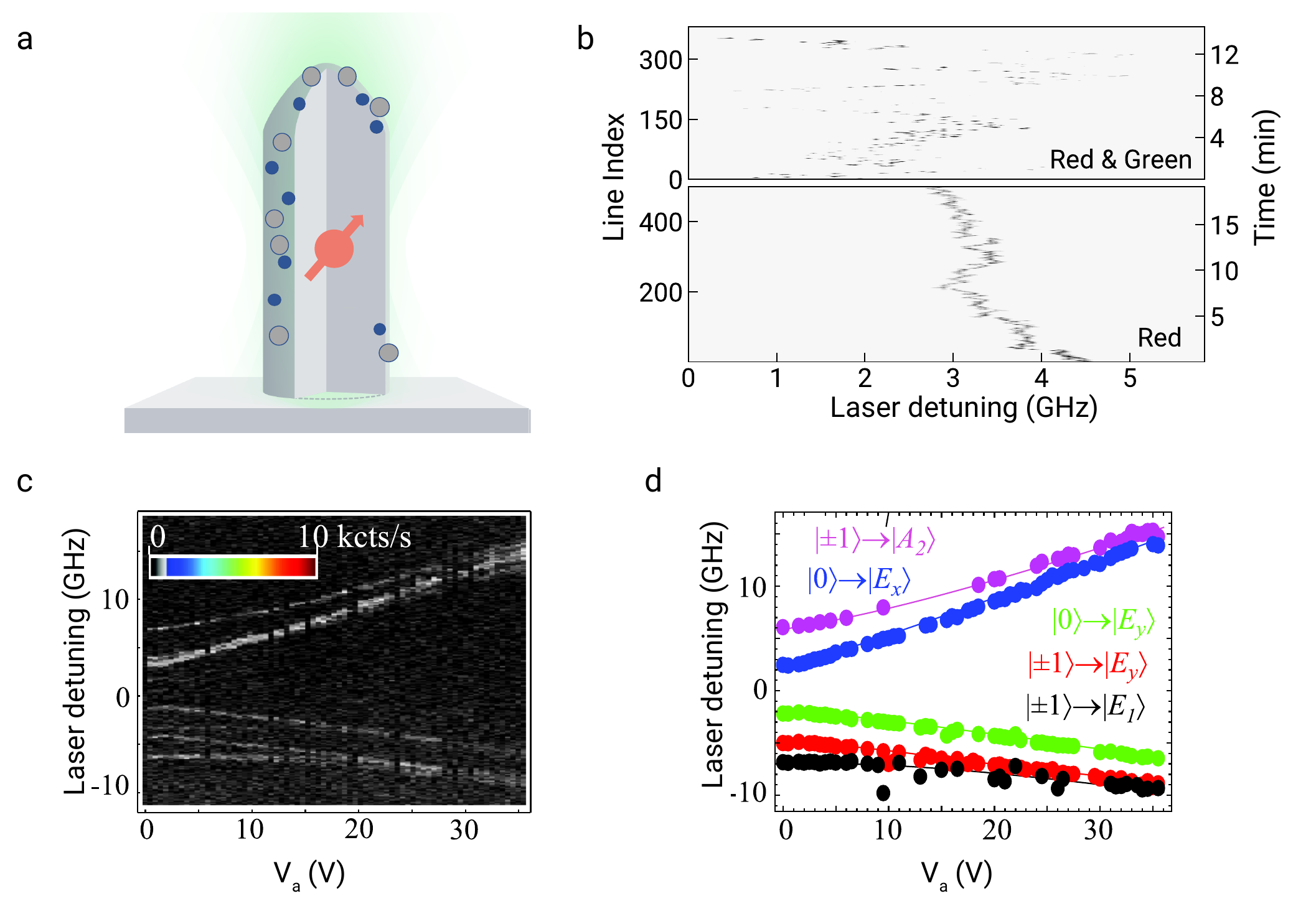}
        \caption{\label{fig:SDfig} Spectral diffusion and spectral control. (a) Schematics, illustrating the process of spectral diffusion of a spin qubit in a diamond nanopillar. At rough surfaces, the density of defects (charge traps and electron donors) is very large. Green laser irradiation leads to photoionization of defects and moving charges, in turn causing fluctuations of the electrostatic environment and spectral diffusion of the ZPL resonances (DC Stark shift). (b) In photoluminescence excitation (PLE) measurements, the spectral dynamics of the ZPL resonance is measured. While additional dim green laser irradiation causes large spectral jumps of several GHz, red laser irradiation only induces slow drifts of the ZPL resonance and spectral stability can be preserved over minutes (no initialization pulses applied). (c) Active tuning of the NV ZPL transition frequencies can be achieved by applying an external voltage. Here, PLE spectra as a function of applied voltage are shown. (d) The ZPL transition frequencies can be modelled with good agreement.
        Figure (b) is adapted from Orphal-Kobin \textit{et al.}~\cite{orphal-kobin_optically_2023} under the terms of the Creative Commons Attribution 4.0 International (CC BY 4.0) license. Copyright 2023, American Physical Society. Figures (c) and (d) are adapted from Acosta \textit{et al.} \cite{AcostaPRL2012} under the terms of the Creative Commons Attribution 3.0 International (CC BY 3.0) license. Copyright 2012, American Physical Society.}
    \end{figure*}

For high-fidelity resonant optical initialization and readout necessary for spin-photon entanglement, as well as to maintain optical coherence of the emitted single photons, spectral stability is a crucial prerequisite. Optical pulses with high energy and large power not only excite or re-initialize the color center, but also lead to ionization of other defect states, i.e., nearby defects in the bulk diamond or defect states on the surface of diamond nanostructures (see \textbf{Figure~\ref{fig:SDfig}}(a)). Random changes in the electrostatic environment over time affect the ZPL transition energy via the DC Stark shift, which in turn leads to spectral diffusion. Depending on the color center's crystallographic geometry and defect environment, an inhomogeneous broadening of the ZPL resonance far exceeding the lifetime-limited linewidth can be observed.

The DC Stark shift contributes directly to the Hamiltonian. It is defined as 
\begin{equation}
    \Delta E_{\rm Stark} =  -\Delta\textbf{d}(\textbf{F})\cdot \textbf{F} = -\Delta dF-\frac{1}{2}\Delta\alpha F^2\,,
\end{equation}
where $\textbf{F}$ is the electric field, and $\Delta d$ and $\Delta \alpha$ represent the field-induced difference of the permanent electric dipole moment and the polarizability between the ground and excited orbital states. The effect of spectral diffusion is in particular detrimental for single photon emission from solid-state quantum systems. 

The NV exhibits a permanent electric dipole moment, which makes it in first-order sensitive to external electric fields and susceptible to spectral diffusion. Detailed theoretical work on the effect of electric fields on the excited state manifold can be found in ~\cite{MazeNJP2011,DohertyNJP2011}. In nanostructures, inhomogeneous linewidths up to several GHz are observed.\cite{WoltersPRL2013} However, the use of native NVs in a nitrogen-rich substrate, advanced fabrication methods, and specific control schemes based on resonant excitation enabled maintenance of spectral stability over minutes and the emission of optically coherent photons on a timescale of one second for NVs in nanostructures (NV-to-surface distance of less than 125\,nm)~\cite{orphal-kobin_optically_2023} (see Figure~\ref{fig:SDfig}(b)).

G4V are inversion-symmetric and are therefore insensitive to first-order DC Stark shifts. Moreover, they even show a suppressed response to second-order effects.\cite{DeSantisPRL2021} (Nearly) transform-limited linewidths were demonstrated for different color centers in several works,\cite{RogersNatComms2014,SipahigilPRL2014,GoerlitzNJP2020,ChenNanoLett2022} also in nanodiamonds and nanostructures.\cite{LiPRApp2016, JantzenNJP2016,EvansPRAppl2016,BhaskarPRL2017,TrusheimPRL2020,NahraAVSQS2021,arjona_martinez_photonic_2022} In a recent work, even at temperatures above 10\,K, transform-limited photon emission was shown for the lead-vacancy color center. A model for decoherence processes due to phonon-induced processes for different G4V color centers was derived.\cite{wang_transform-limited_2023-1}

Especially for quantum networks, indistinguishable photons are required for quantum interference, and in turn, for spin-spin entanglement generation. Usually, in a Hong-Ou-Mandel-type (HOM) setup,\cite{Hong1987} two-photon interference on a beam splitter, considered as a linear optical gate, is used for Bell-state measurements. Photonic indistinguishability describing photons with an equal quantum state relates to all optical degrees of freedom: frequency, polarization, temporal mode, and spatial mode. The HOM visibility, often quantified by the coherence time of the emitters, is a direct measure of photon indistinguishability and determines the entanglement generation fidelity. Theoretical predictions for the visibility and the output Bell-state fidelity set by the single-photon characteristics of different emitters are analyzed in detail in ref.~\cite{kambs_limitations_2018}. In two-photon quantum interference experiments with color centers in diamond, often the visibility is determined by measuring the zero-delay second-order intensity correlation function for aligned $g_{\parallel}^{(2)}(0)$ and orthogonal $g_{\perp}^{(2)}(0)$ polarization of the input modes
\begin{equation}
    V =  1-\frac{g_{\parallel}^{(2)}(0)}{g_{\perp}^{(2)}(0)}\,.
\end{equation}
For NVs in microstructures for the first time in 2012, a time-resolved two-photon interference contrast of $66\%$ has been demonstrated\cite{BernienPRL2012} and later on even a visibility of up to $90\%$, applying a pulse train of 10 optical $\pi$-pulses that excite both emitters.\cite{HumphreysNature2018} For the SiV in bulk, a visibility of $72\%$~\cite{SipahigilPRL2014} and for the SnV in a single-mode waveguide a visibility of $63\%$~\cite{arjona_martinez_photonic_2022} was reported. Whereas in experiment, the visibility is not only limited by non-perfect indistinguishability due to spectral broadening, but additionally by contributions from a finite detector time resolution and background events.

Measures that are taken to reduce spectral diffusion, are the improvement of color center formation methods,\cite{vanDamPRB2019,RugarACSNanoLett2020} i.e. using preferentially in-grown defects or only shallow ion implantation and over-growth, as well as high-pressure high-temperature annealing to reduce the number of vacancies in the diamond lattice (see Section~\ref{sec:formation}). Moreover, micro- and nanofabrication methods are optimized to reduce the number of surface defects.\cite{LekaviciusJApplPhys2019, RufNanoLett2019} Certain surface terminations preserve the spin coherence for shallow color centers~\cite{SangtawesinPRX2019} and might also improve optical coherence. Active approaches to mitigate spectral diffusion are adjusted excitation schemes~\cite{orphal-kobin_optically_2023,Gorlitznpj2022} and active feedback schemes.\cite{AcostaPRL2012} Furthermore, methods based on pulsed optical control protocols have been proposed for suppressing spectral diffusion~\cite{FotsoPRL2016} and even demonstrated for spectral shaping, for example, using SiC.\cite{lukin_spectrally_2020}

The sensitivity of color centers to external electric fields~\cite{AcostaPRL2012, AghaeimeibodiPRAppl2021} and to strain~\cite{MartinSanchezSemiSciTech2018,MeesalaPRB2018,MachielsePRX2019} can be used for active tuning of the emitter's resonances up to several GHz (see Figure~\ref{fig:SDfig}(c) and (d), Section \ref{sec: MechCtrl}). 

\subsection{Optical micro- and nanostructures for controlled light-matter interaction }\label{sec:light-matter}

In the past years, a multitude of diamond nanostructures evolved.\cite{SchroederJOSAB2016} Often, these structures aim for better photon collection efficiencies, i.e. lower photon losses, which are currently an obstacle not only for fiber-based quantum communication networks.\cite{BorregaardPRX2019}
Amongst others, bullseye gratings,\cite{LiNanoLett2015} parabolic reflectors \cite{WanNanoLett2018} as well as inverted nanocones \cite{torun_optimized_2021} have been investigated.

Nanocavities denote a special class of diamond nanostructures since they do not only affect the embedded color center's spatial emission characteristics, but also the interaction strength between photons and the color centers themselves.\cite{Haroche1989, Janitz2020}
In general, the interaction cross section between light and single (artificial) atoms is weak.
It can be increased significantly employing an optical cavity which surrounds the atom.\cite{Reiserer2015}
Thus, embedding a diamond color center into a photonic crystal (PhC) cavity etched into the diamond substrate, enhances on the one hand single-photon generation processes and allows for optical nonlinearities (refer to Section~\ref{sec:optctrl_cavity_spin_manipulation}) on the other hand.

In the simplest case, an atom with a ground state $\ket{g}$ and an excited state $\ket{e}$ is coupled to a cavity with a quality factor $Q=\omega_\mathrm{c}/\left(2\kappa\right)$ derived from the cavity's resonance frequency $\omega_\mathrm{c}$ and its decay rate $\kappa$.
Furthermore, the cavity possesses a mode volume
\begin{align}
    V = \frac{\bigintss_{\vec{r}} \epsilon\!\left(\vec{r}\right)\left|\vec{E}\!\left(\vec{r}\right)\right|^2 \mathrm{d}^3 \vec{r}}{\max\!\left[\epsilon\!\left(\vec{r}\right)\left|\vec{E}\!\left(\vec{r}\right)\right|^2\right]}
\end{align}
defined as an integral over the cavity mode's spatial electric field intensity distribution $\left|\vec{E}\!\left(\vec{r}\right)\right|^2$ multiplied by the relative permittivity $\epsilon$ at position $\vec{r}$ and normalized to the maximal field energy.
Altogether with the vacuum permittivity $\epsilon_0$, the reduced Planck constant $\hbar$, and the respective electric transition dipole matrix element $\mu_\mathrm{ge}$, the single-photon coupling constant becomes \cite{Pinkse2003, Reiserer2015}
\begin{align}
    \label{eq:cavity_g}
    g = \sqrt{\frac{\mu_\mathrm{ge}^2 \omega_\mathrm{c}}{2\hbar\epsilon_0 V}} \,.
\end{align}
With the atom's decay rate $\gamma$, the emitter-cavity cooperativity is $C=g^2/\left(2\kappa\gamma\right)$.
If $C\gg 1$, the single cavity-embedded atom is able to interact nonlinearly with light \cite{Reiserer2015}.
Two regimes fulfill the condition $C\gg 1$.
Firstly, the weakly-coupled Purcell regime identified by $\kappa/2\gg g\gg\gamma$ leads to Purcell enhancement:
the cavity-incorporated emitter's emission rate is risen by the Purcell factor $F_\mathrm{P}$ compared to its bulk emission rate.\cite{Purcell1946}
This also increases the emitter's homogeneous linewidth, which is advantageous to approach a spectral overlap between multiple distinct emitters (see Section~\ref{sec:spectral_control}).
Likewise, the emitter's Purcell factor-dependent Debye-Waller factor $DW$ rises.
Secondly, the strongly-coupled regime characterized by $g\gg\gamma,\kappa$ leads to new effects facilitating optical spin control and readout as well as single-photon nonlinear processes as detailed in the next section.

Nowadays, there are several realizations of diamond color centers coupled to PhC cavities:
NV centers have been coupled to two-dimensional \cite{FaraonPRL2012, RiedrichMoeller2015} and one-dimensional nanobeam \cite{LiNatComms2015, MouradianAPL2017} cavities with rectangular cross-sections.
For SiV centers, a two-dimensional cavity \cite{RiedrichMoellerNanoLett2014} and one-dimensional cavities with a triangular cross-section \cite{Zhang2018, NguyenPRL2019, Knall2022} exist.
Recently, one-dimensional nanobeam cavities with a rectangular cross-section containing SnV centers have been fabricated.\cite{RugarPRX2021, KurumaAPL2021}

\begin{figure*} 
    \centering 
    \includegraphics[width=\textwidth]{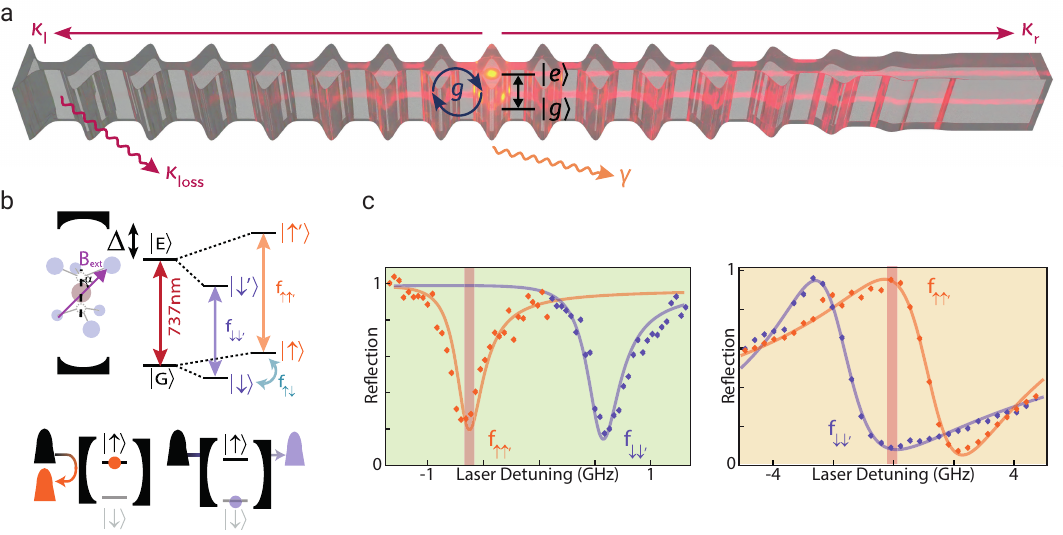}
    \caption{\label{fig:cavity_qed}
         Jaynes–Cummings model and spin-dependent photon reflection.
         (a) Model of a single (artificial) atom (orange sphere) coupled to a PhC cavity (gray corrugated structure) leading to the Jaynes–Cummings Hamiltonian in cavity quantum electrodynamics.
         The embedded atom possesses a ground state $\ket{g}$ and an excited state $\ket{e}$ with a decay rate $\gamma$.
         It couples to the (asymmetric) cavity with a rate determined by the single-photon coupling constant $g$.
         The total cavity decay rate $\kappa$ consists of the left and right (incoupling) cavity mirror's respective decay rates $\kappa_\mathrm{l}$ and $\kappa_\mathrm{r}$ as well as of the cavity scattering losses $\kappa_\mathrm{loss}$.
         Thus, it becomes $\kappa=\kappa_\mathrm{l}+\kappa_\mathrm{r}+\kappa_\mathrm{loss}$.
         (b) Under an atom-cavity detuning $\Delta$, the SiV's $\SI{737}{\nano\meter}$ optical transition (red) couples to a PhC cavity.
         An external magnetic field $B_\mathrm{ext}$ splits the SiV's spin-conserving transitions (orange and blue) allowing for spin-selective reflection at the cavity input interface: photons at a fixed frequency are reflected (transmitted) if the SiV is in spin state $\ket{\uparrow}$ $\left(\ket{\downarrow}\right)$.
         (c) Cavity reflectivity $R\!\left(\Delta_\mathrm{c}\right)$ depending on the embedded SiV's spin state for a large (small) SiV-cavity detuning depicted in the left (right) panel.
         With a large detuning $\Delta$, the transitions $f_\mathrm{\uparrow\uparrow^{'}}$ and $f_\mathrm{\downarrow\downarrow^{'}}$ can be clearly distinguished.
         However, a small detuning $\Delta$ enables the highest contrast (red shaded areas) in the spin-dependent cavity reflectivity.
         Figure (b) is reprinted with permission from Nguyen \textit{et al}.\cite{NguyenPRL2019} Copyright 2019 American Physical Society. Figure (c) is reprinted with permission from Nguyen \textit{et al}. \cite{NguyenPRB2019} Copyright 2019 American Physical Society.
    }
 \end{figure*}

All of the aforementioned cavity designs rely on holes or elliptic features.
To extract part of the light $\kappa_\mathrm{out}$ stored in a PhC cavity without introducing scattering losses $\kappa_\mathrm{loss}$, tapered cavity-waveguide interfaces with slowly vanishing cavity features are employed \cite{Knall2022} (see \textbf{Figure~\ref{fig:cavity_qed}}(a)).
Such interfaces might suppress scattering losses even better maintaining the cavity performance common for hole-based designs by replacing the holes with corrugation-based 'Sawfish' designs.\cite{Bopp2022, Pregnolato2023}
Considering the cavity's total decay rate $\kappa = \kappa_\mathrm{out} + \kappa_\mathrm{loss}$, the cavity-waveguide coupling efficiency $\beta_\mathrm{WG}=\kappa_\mathrm{out}/\kappa$, and the probability to emit a photon into the cavity mode $\beta_\mathrm{C}=2C/\left( 1+2C\right)$, the overall probability $\eta$ to firstly emit a ZPL photon into the cavity and to secondly transfer it to the attached waveguide becomes $\eta=DW\times\beta_\mathrm{C}\beta_\mathrm{WG}$.\cite{Reiserer2015}

To eventually interface diamond waveguides with optical fibers by evanescent coupling, a tapered fiber tip is brought in close contact with a diamond waveguide which narrows down adiabatically.\cite{Tiecke2015}
Recent works apply these fiber-optical interfaces due to their near-unity coupling efficiencies.\cite{BhaskarPRL2017, NguyenPRL2019, NguyenPRB2019, Knall2022, parker_diamond_2023}
Similar adiabatic photonic interfaces between diamond and silicon or aluminium nitride waveguides allow for hybrid integration of diamond waveguides \cite{MouradianPRX2015} and microchiplets \cite{wan_large-scale_2020} by pick-and-place.
Hybrid integration methods, more scalable than pick-and-place, comprise wafer bonding and transfer printing.\cite{elshaari_hybrid_2020, kim_hybrid_2020}

\subsection{Cavity-mediated spin-photon interaction}
\label{sec:optctrl_cavity_spin_manipulation}

(Artificial) atoms with an optical transition of frequency $\omega_\mathrm{a}$ between a ground $\ket{g}$ and an excited $\ket{e}$ state coupled to (PhC) cavities do not only offer increased emission rates and an improved photon collection efficiency.
Under the term `cavity quantum electrodynamics', a multitude of effects occurring for cavity-coupled atoms has been studied.\cite{Agarwal2012}
These effects involve the ability of a cavity to reflect single photons impinging on its mirrors depending on whether an atom is coupled to the cavity or whether the cavity is empty. This effect can be a fundamental ingredient for spin-photon entanglement creation.

Off-resonance, an empty optical cavity reflects photons with unity probability.
However, if such a cavity is tuned to resonance, a Lorentzian dip with a full width at half maximum of $2\kappa$ appears in its reflectivity.
With an atom coupled to the cavity, the reflectivity changes dramatically -- even in the Purcell regime.
Applying input-output theory to the Jaynes–Cummings Hamiltonian allows to derive an expression for the frequency-dependent amplitude reflection coefficient $r\!\left(\omega\right)$ of a coupled atom-cavity system.\cite{Reiserer2015, Hu2008}
This coefficient relates to the reflectivity via $R\!\left(\omega\right)= \left|r\!\left(\omega\right)\right|^2$.
With the cavity (atomic) detuning $\Delta_{\mathrm{c}\left(\mathrm{a}\right)} \equiv \omega - \omega_{\mathrm{c}\left(\mathrm{a}\right)}$ from the probe frequency $\omega$, the cavity's strong mirror decay rate $\kappa_\mathrm{l}$, its weak (incoupling) mirror decay rate $\kappa_\mathrm{r}$, and the atom's excited state decay rate $\gamma$ (see Figure~\ref{fig:cavity_qed}(a)), it is
\begin{align}
    \label{eq:cavity_qed_r}
    r\!\left(\omega\right) = 1 - \frac{2 \kappa_\mathrm{r} \left(\mathrm{i} \Delta_\mathrm{a} + \gamma\right)}{\left(\mathrm{i} \Delta_\mathrm{c} + \kappa\right) \left(\mathrm{i} \Delta_\mathrm{a} + \gamma\right) + g^2}\,.
\end{align}

For an atom coupled to a cavity under zero atom-cavity detuning $\Delta\equiv \left|\omega_\mathrm{c}-\omega_\mathrm{a}\right| = 0$, the Lorentzian dip in the cavity's reflectivity with a minimum at $\Delta_\mathrm{c}=0$ becomes replaced by two dips offset from the cavity resonance by the coupling constant $\pm g$.\cite{Reiserer2015}
If only an optical transition related to one of the cavity-embedded atom's spin states couples to the cavity, this effect enables mapping time-bin photonic qubits to the atom's spin degree of freedom by rendering the atom-cavity coupling spin-dependent \cite{NguyenPRB2019} (see Figure~\ref{fig:cavity_qed}(b)).
In contrast to empty cavities ($g=0$), the phase $\phi\!\left(\omega\right)=\arg\!\left[r\!\left(\omega\right)\right]$ of the light reflected off the atom-coupled cavity does not undergo a phase shift around the cavity resonance $\Delta_\mathrm{c}=0$.
The phase difference between photons reflected off an empty and an atom-coupled cavity can be engineered to be $\mathrm{\pi}$.\cite{Reiserer2015, Hu2008}
Particularly for a spin-dependent atom-cavity coupling, this enables a further polarization-based route towards entanglement generation.\cite{ChenNPJQI2021}

At a small atom-cavity detuning $\Delta < \kappa$, $R\!\left(\omega\right)$ describes a Fano line shape, which yields the highest contrast in the cavity reflectivity depending on the atom-cavity coupling.
As soon as the detuning rises further ($\Delta > \kappa$), the reflectivity contrast decreases, but the atom-cavity system's resonances become narrower.
For a cavity-coupled SiV, this regime leads to highly spin-conserving optical transitions, and thus allows for efficient single-shot readout of the SiV's spin configuration \cite{NguyenPRB2019} (see Figure~\ref{fig:cavity_qed}(c)).

Employing diamond color centers as artificial atoms integrated into diamond nanocavities and utilizing their spin-dependent optical transitions to spin-selectively couple to the cavity, a couple of pioneering studies establish cavity-mediated spin readout and control schemes.
These schemes are exploited amongst others for spin-photon entanglement generation. In 2018, two SiV centers were coupled to a diamond nanocavity and the cavity's reflectivity was measured dependent on the spin state of one of the two SiV centers.\cite{EvansScience2018}
Furthermore, controlling the SiV spin states, photon-mediated interactions between the two SiV centers could be deterministically switched on and off.
In a follow-up work, a cavity-embedded SiV center was used to generate Bell states between the SiV spin and a photon being reflected off the cavity, again depending on the SiV spin state.\cite{NguyenPRB2019}
Also, the usability of such a system as a node within a quantum network to implement a quantum repeater was examined.\cite{NguyenPRL2019}
Rising the same system's cooperativity to $C=\num{105(11)}$ recently allowed the demonstration of non-destructive single-shot cavity reflection-based readout of the SiV spin state with a fidelity of about $\num{99.98}\,\%$.\cite{BhaskarNature2020}
To map polarization-encoded photonic qubits to the spin degree of freedom of a solid-state artificial atom, the PEPSI scheme was proposed.\cite{ChenNPJQI2021}
In a first implementation of a single-photon nonlinearity based on a waveguide-coupled SnV, a spin-gated $\num{11(1)}\,\%$ decrease in the reflectivity was demonstrated.\cite{parker_diamond_2023}
Moreover, recently, a two-node quantum network in the Boston urban area could be realized, facilitated by reflection-based spin-photon gates employing cavity-embedded SiV centers and time-bin encoding.\cite{knaut_entanglement_2023}

\section{Mechanical control} \label{sec: MechCtrl}

\begin{figure*}[ht!]
        \centering
        \includegraphics[width=.7\textwidth]{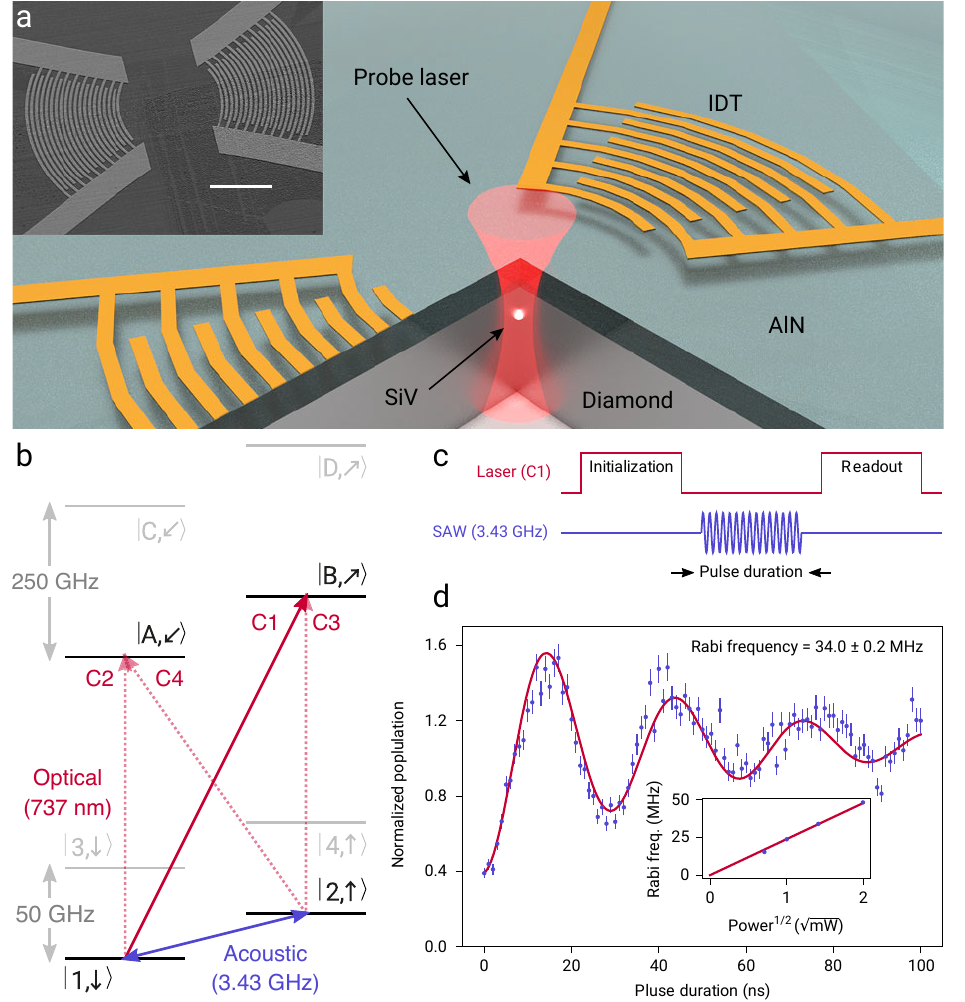}
        \caption{\label{fig:mechcontrol} 
        Overview of the diamond-based surface accoustic wave (SAW) device used in \cite{maity_coherent_2020}. (a) A microwave signal, when applied to one of the interdigital transducers (IDTs), leads to the generation of acoustic waves, due to the piezoelectric effect in aluminum nitride (AlN). A focused laser beam is used to interrogate silicon-vacancies (SiVs). (Inset) 
        The scanning electron microscope (SEM) image shows a pair of microwave transducers. The scale bar shows 20\,µm. (b) The electronic structure of the SiV when subjected to an external magnetic field. The solid red arrow points to the optical transition used for the spin's initialization and reading, while the dashed red arrows show other optical transitions. The blue arrow marks the acoustic transition between the spin qubit's two levels. 
         (c) A pulse sequence for coherently driving the SiV involving an acoustic pulse at 3.43 GHz,
        which matches the SiV spin transition frequency. The SAW pulse is produced with a peak microwave input of 2 mW. (d) Observation of the normalized population in the $|1\downarrow\rangle$ state 
         as the duration of the acoustic pulse is varied. A theoretical model fit is depicted in red. (Inset) The Rabi frequency as a function of the square root of the peak microwave input power shows 
         the anticipated linear trend. The Rabi frequency errors are around 0.1 MHz.   
        Figures (a)-(d) are rearranged and adapted from Maity \textit{et al.} \cite{maity_coherent_2020} under the terms of the Creative Commons Attribution 4.0 International (CC BY 4.0) license. Copyright 2020, Nature Publishing Group. 
        }
    \end{figure*}

The interplay between spin and the crystal strain offers another avenue for coherent control of spin qubits in diamond. 
The key mechanism behind the spin-phonon interaction is the coupling of the color center to strain,\cite{teissier_strain_2014, maity_coherent_2020} or in other words: to a displacement of the carbon atoms in the color center's environment. The displacement directly interacts with the electronic degree of freedom through a distortion of the orbitals that are populated by the color center's electrons. In the presence of a magnetic field, dynamic strain caused by either bulk oscillations, i.e. phonons, or macroscopic oscillations of the diamond system, can then directly influence the electronic spin degree of freedom as shown in \textbf{Figure~\ref{fig:mechcontrol}}.

To leverage the interaction with the mechanical degree of freedom many approaches have been explored.
One such avenue involves the harnessing of surface acoustic waves (SAWs),\cite{schuetz_universal_2015, golter_optomechanical_2016, maity_coherent_2020} which interact with shallow implanted color centers. An example of a SAW device for control of an SiV is shown in Figure~\ref{fig:mechcontrol}a.  The utilization of mechanical oscillators, such as bulk mechanical oscillators,\cite{macquarrie_mechanical_2013, macquarrie_coherent_2015} an oscillating cantilever,\cite{teissier_strain_2014, barfuss_strong_2015, clark_nanoelectromechanical_2023} and diamond microdisk resonators \cite{shandilya_optomechanical_2021} have also been implemented.
A more extensive description of mechanical spin interfaces and their properties can be found in the topical reviews.\cite{lee_topical_2017, wang_coupling_2020}
Mechanical control has also been achieved by magnetically linking the oscillation of a mechanical resonator to NVs.\cite{hong_coherent_2012, pigeau_observation_2015} For this review, we categorize it under microwave control since the resonator's motion mainly exposes the color center to a time-dependent magnetic field.

The interest for exploring mechanical interactions with an optically active spin system are wide ranging. It starts with understanding the fundamentals of the spin-phonon coupling mechanism with the electronic spin of a color center, which has direct implications for its coherence properties,\cite{JahnkeNJP2015, harris_coherence_2023} their coherent optical efficiency,\cite{zhang_vibrational_2011, alkauskasNPJ2014} and ensemble properties that can be greatly affected by a dynamic strain environment.\cite{bennett_phonon-induced_2013, li_preparing_2020} Furthermore, the mechanical interaction can be used to couple a macroscopic mechanical degree of freedom to a quantum degree of freedom, allowing for the study of quantum mechanical oscillators and for creating non-classical states of these hybrid systems.\cite{arcizet_single_2011} Another interesting application for spin-mechanical hybrid systems lies in the transduction of microwaves to the optical regime for coupling superconducting qubits in a quantum network with the goal of scaling quantum computers.\cite{rabl_quantum_2010, li_hybrid_2015, kim_diamond_2023} Phononic interactions can also be used to create phononic quantum networks connecting spins.\cite{habraken_continuous_2012, lemonde_phonon_2018} In recent research,\cite{clark_nanoelectromechanical_2023} a device named "Strain-Transduction by Resonantly Actuated Integrated Nanoelectromechanical Systems" (STRAINEMS) was introduced. Its primary purpose was to show how strain can be adjusted and controlled at the chip level for color centers. This work serves as a foundational step towards establishing that mechanical control can be scaled up for quantum information processors. 

The application that will take center stage here is the coherent control of spin defects in diamond.\cite{macquarrie_mechanical_2013, maity_coherent_2020} Establishing such control will have larger implications, due to the spin hyperfine interaction used for controlling e.g. ${}^{13}$C spins in the environment of the electronic spin.\cite{maity_mechanical_2022} 

\subsection{Mechanical control of G4Vs}

Because the spin-strain coupling strength is quite large for G4Vs ($1.5$ PHz/strain, SiV \cite{maity_coherent_2020}) compared to the NV ($5.46 - 19.63$ GHz/strain \cite{teissier_strain_2014}),  mechanical control is an interesting avenue for efficient control of the G4V, reducing for example heating during control sequences. The mechanical control is facilitated through the interaction of a color center with strain. In \cite{maity_coherent_2020} the authors outline the coupling of a spin to a dynamical strain field to the SiV, which can be generalized to other G4Vs. The Hamiltonian that best captures the spin-strain interaction is 
\begin{equation}
    H_{\rm G4V} = H_{\rm SO} + H_{\rm Z} + H_{\rm strain}\,,
\end{equation}
where $H_{\rm SO}$ is the spin-orbit interaction, $H_{\rm strain}$ is the strain interaction, and $H_{\rm Z}$ the axial Zeeman Hamiltonian which describes the interaction with a magnetic field. In the basis $|E_{\rm gx},\downarrow\rangle, |E_{\rm gy}, \downarrow\rangle, |E_{\rm gx},\uparrow\rangle, |E_{\rm gy}, \uparrow\rangle$ -- a direct product of the orbital degrees of freedom  ($E_{\rm gx}, E_{\rm gy}$) and the spin degrees of freedom ($\downarrow, \uparrow$) -- the spin-orbit and Zeeman term are given by \cite{hepp_electronic_2014}
\begin{align}
    H_{\rm SO} & = \frac{\lambda}{2}
    \sigma_y
    \otimes 
    \sigma_z ~,
    \\
    H_{\rm Z} & = 
    \frac{\gamma_s}{2} B_z 
    \mathbbm{1}
    \otimes
    \sigma_z
    +
    \frac{\gamma_s}{2}(B_x\mathbbm{1}\otimes \sigma_x + 
    B_y \mathbbm{1}\otimes \sigma_y)\,,
\end{align}
where $\lambda$ is the the magnitude of the spin-orbit coupling, $B_i$ are the components of the magnetic field in the internal reference frame of the G4V,\cite{hepp_electronic_2014} $\sigma_i$ are the Pauli matrices, and $\gamma_s$ is the gyromagnetic ratio.  
For the Zeeman term, the coupling between different orbitals is neglected because of the quenching of the orbital interaction. The electronic level structure of the SiV is shown in Fig.~\ref{fig:mechcontrol}b, with the respective optically and acoustically coupled levels. 

The unperturbed G4Vs are part of the D$_{\rm 3d}$ symmetry group, coupling to phonons with A$_{\rm 1g}$ and E$_{\rm g}$ symmetry:
\begin{equation}
    H_{\rm strain} = H_{\rm strain}^{\rm A_{\rm 1g}} + H_{\rm strain}^{\rm E_{\rm g}}\,.
\end{equation}
The A$_{\rm 1g}$ representation of the strain tensor \cite{hepp_electronic_2014, maity_coherent_2020} is given by
\begin{equation}
    H^{\rm A_{\rm 1g}}_{\rm strain} = \alpha
    \mathbbm{1}
    \otimes 
    \mathbbm{1}\,,
\end{equation}
where $\alpha = t_{\perp} (\epsilon_{xx} + \epsilon_{yy}) + t_\parallel \epsilon_{yy}$, $\epsilon_{ik}$ is a component of the strain tensor $\epsilon$ and $t_{\perp/\parallel}$ can be determined experimentally. $H^{\rm A_{1g}}_{\rm strain}$ induces a global shift of the energy values and do not couple the spins. The E$_{\rm g}$ components, however, can induce spin mixing, in the presence of an off-axis magnetic field. The representation of the E$_{\rm g}$ component of the strain is
\begin{equation}
    H^{\rm E_{\rm g}}_{\rm strain} = 
    (-\beta \sigma_z + \gamma \sigma_x) \otimes 
    \mathbbm{1}\,,
\end{equation}
where $\beta = d ( \epsilon_{xx} - \epsilon_{yy})  + f \epsilon_{zx}$ and $\gamma = -2d \epsilon_{xy} + f \epsilon_{yz}$. Again, $d, f$ are scaling factors, which can be determined empirically. 
The magnitude of the spin-strain response for an slightly off-axis magnetic field can be determined from $\Omega = \langle E_{g+}, \downarrow' | H_{\rm strain} |E_{g-}, \uparrow'\rangle$, where $|E_{g+}, \downarrow'\rangle$ and $|E_{g-},\uparrow'\rangle$ are the two lowest energy eigenstates of $H_{\rm SO} + H_{\rm Z}$. It is assumed that $|B_x|, |B_y| \ll |B_z|$ acts as a perturbation. In \cite{maity_coherent_2020} the authors find
\begin{align}
    &|E_{g+}, \downarrow' \rangle \approx |E_{g+}, \downarrow\rangle + \frac{\gamma_s}{2\lambda}(B_x + {\rm i}B_y)|E_{g+},\uparrow\rangle \\
    &|E_{g-}, \uparrow' \rangle \approx |E_{g-}, \uparrow\rangle + \frac{\gamma_s}{2\lambda}(B_x - {\rm i}B_y)|E_{g-},\downarrow\rangle\,,
\end{align}
where $|E_{g\pm} \rangle = |E_{gx}\rangle \pm {\rm i}|E_{gy}\rangle$
so that 
\begin{equation}
    \Omega = -\frac{\gamma_s}{\lambda}(\beta + {\rm i}\gamma)(B_x - {\rm i}B_y)\,.
\end{equation}
The spin-strain Hamiltonian in the new basis $|E_{g+}, \downarrow'\rangle, |E_{g-},\uparrow'\rangle$ then is 
\begin{equation}
    H = 
    -\frac{\epsilon}{2}\sigma_z
    + {\rm Re}[\Omega]\sigma_x - {\rm Im}[\Omega]\sigma_y\,,
\end{equation}
where $\epsilon = \gamma_s B_z /2$ are the eigenenergies of the non-interacting system. 
A time-dependent strain field with $\beta(t)$ and $\gamma(t)$ at the position of the color center, for example caused by surface acoustic waves can then be used to induce rotations on the spin qubit's Bloch sphere, which are mediated by the generalized Rabi frequency $\Omega(t)$. In summary, mechanical control of the spin qubit of a G4V in the presence of a slightly off-axis magnetic field is possible, if a time-dependent strain field interacts with the G4V. In Fig.~\ref{fig:mechcontrol}c a pulse sequence is shown for the initialization of the spin state followed by an acoustic control and readout pulse. The obtained Rabi oscillations as a function of the acoustic pulse duration can be seen in Fig.~\ref{fig:mechcontrol}d. 

The hyperfine interaction of the electronic spin with the nuclear spins of ${}^{13}$C in the environment of the G4V, can be leveraged to extend the mechanical control to the nuclear spins.\cite{maity_mechanical_2022}  

\subsection{Mechanical control of the NV}

Similar to the G4Vs, the negatively charged NV spin can interact with a strain field. In contrast to the G4V, the NV is a spin-1 system and has C$_{\rm 3v}$ symmetry, which changes the structure of the interaction and therefore the details of the mechanical control:\cite{teissier_strain_2014}
\begin{equation}
H / h=\underbrace{D_0 S_z^2}_{H_0}+\underbrace{\gamma_{\mathrm{NV}} \vec{S} \cdot \vec{B}}_{H_{\text {Zeeman }}}+\underbrace{d_{\|} \epsilon_z S_z^2-d_{\perp} / 2\left[\epsilon_{+} S_{+}^2+\epsilon_{-} S_{-}^2\right]}_{H_{\text {strain }}}\,,    
\end{equation}
where $D_0 = 2.87$~GHz is the zero-field splitting, $\gamma_{\rm NV}= 2.8$~MHz/G the NVs gyromagnetic ratio, $h$ Planck's constant, $\vec B$ the external magnetic field, and $\vec S$ is the NV electron spin operator with components $S_x$, $S_y$ and $S_z$, and $S_\pm = S_x \pm \i S_y$. 
The Hamiltonian $H_{\rm strain}$ describes the coupling of the NV spin to lattice strain $\epsilon_j$ along coordinate $j \in \{[100], [010], [001]\}$, where $[ijk]$ are the lattice directions of the fcc-diamond lattice and $\epsilon_\pm = -\epsilon_y \pm \i\epsilon_x$. 
The strain coupling constants corresponding to strain longitudinal and transverse to the NV axis are denoted by $d_\parallel = 5.46$ GHz/strain and $d_\perp = 19.63$ GHz/strain.

The influence of $\epsilon_z$ is a modification of $D_0$ and thus, only alters the energy difference between the states $|\pm 1\rangle$ and $|0\rangle$ (eigenstates of $\vec S$). On the contrary, with $\epsilon_{x,y} \neq 0$, the states $|+1\rangle$ and $|-1\rangle$ can mix and are therefore amenable to mechanical control.

An oscillating $B_\perp$ field can drive transitions between $|0\rangle$ and $|\pm 1\rangle$, whereas $d_\perp$ can directly couple $|+1\rangle$ and $|-1\rangle$, which are not directly coupled by $B_\perp$. This allows for establishing direct transitions between all levels of the NV's ground state manifold. 

\section{Perspectives}\label{sec:perspectives}

High fidelity coherent quantum control is a necessary prerequisite for almost all quantum applications, which can be roughly divided into two broad categories: applications that do not require entanglement and those that do. Sensing, for example, does not necessarily require entanglement, but can greatly benefit from it.\cite{degenRevModPhys2017} For most future quantum information applications, however, such as wide area quantum networks and quantum computing, entanglement is the key enabling quantum phenomenon. Here we seek to provide a non-extensive overview of some of the potential applications for which color centers may produce the required entanglement either by exploiting interactions with nuclear spins, electronic spins, phonons as well as photons. 

\begin{figure*}[ht!]
    \centering
    \includegraphics[width=.9\textwidth]{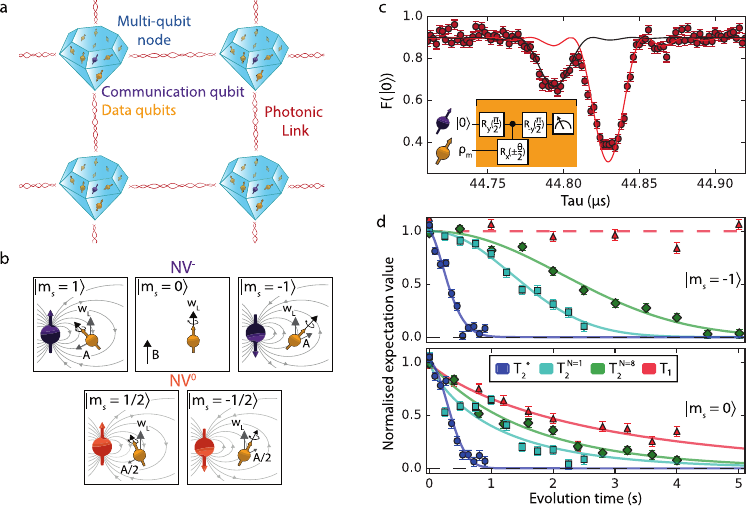}
    \caption{\label{fig:quantum_network_architecture}
        Distributed quantum information processing with diamond-based spins: (a) Formation of multi-qubit nodes utilizing electron spins for 'communication' (depicted in purple) and nuclear spins for 'data' (shown in yellow). 
        These allow for complete control, with the electron spins' ability to coherently interact with photons facilitating the heralded creation of photonic entanglement across distant nodes. 
        The nuclear spins contribute to the prolonged preservation and manipulation of the quantum information. 
        (b) Interaction between electron and nuclear spins: The nuclear spins' rotation at the Larmor frequency around an external magnetic field ${\rm \bf B}$ is influenced by the electron spin's presence. 
        Variations in the rotation frequency and axis occur based on the charge and spin states of the NV, 
        which include NV$^-$ (in purple, $m_s = \{-1, 0, +1\})$ and NV$^0$ (in orange, 
        $m_s = \{-1/2, +1/2\}$). 
        (c) NV centre's dynamical decoupling spectroscopy: 
        A specific resonance at 44.832 $\mu$s (highlighted in red) indicates a solitary ${}^{13}$C spin, characterized by hyperfine parameters $A_\parallel = 2\pi\times80(1)$ Hz and $A_\perp = 2\pi\times271(4)$ Hz. 
        Another resonance at 44.794 $\mu$s (in black) relates to the collective ${}^{13}$C spin bath. 
        The solid lines represent a theoretical model (details in Supplementary Note of \cite{bradley_robust_2022}). 
        (d) Nuclear spin's intrinsic decoherence timescales for various electron states: Solid lines indicate model fits, while dashed lines offer a visual guide. 
        'N' signifies the count of spin-echo pulses. 
        Error bars represent one standard deviation. 
        Figure and caption adapted from Bradley \textit{et al.} \cite{bradley_robust_2022} under the terms of the Creative Commons Attribution 4.0 International (CC BY 4.0) license. Copyright 2022, Nature Publishing Group.
        }
\end{figure*}

As mentioned in previous sections, color centers such as the NV \cite{PompiliScience2021} and SiV \cite{BhaskarNature2020} have already made an impact in the realm of quantum networks, where one of the key challenges is the distribution of entanglement across various nodes in the network. An implementation of a network with diamond spins and the NV is shown in \textbf{Figure~\ref{fig:quantum_network_architecture}} which was adapted from \cite{bradley_robust_2022}.
Quantum networks have wide ranging applications, some of them being the well known quantum key distribution, blind quantum computing, worldwide timekeeping, navigation and sensing, fundamental test of physics as well as the scaling of quantum computers (see e.g. \cite{simon_towards_2017} and references therein). 

For quantum network applications, spin-photon entanglement \cite{ToganNature2010} is often the most intense focus of research efforts: 
The stationary electronic spin hosted by a color center either directly acts as a memory or as a quantum bus between a longer lived memories made up of nuclear spins.\cite{PompiliScience2021} 
An excellent overview of methods for establishing entanglement between stationary qubits and photons can, for example, be found in the tutorial \cite{beukers_tutorial_2023}. Here, the authors provide a high level overview over various building blocks for entanglement protocols. 

Once a flying qubit is entangled with the spin-qubit, spin-spin entanglement is the next logical step, which was demonstrated in a range of seminal works \cite{BernienNature2013, HensenNature2015, KalbScience2017, HumphreysNature2018, PompiliScience2021} with NVs at the core of the spin-photon interface. 
Similarly, the SiV has made great strides towards quantum networks, from the first demonstrations of a heralded entanglement scheme,\cite{NguyenPRB2019} memory-enhanced quantum communication \cite{BhaskarNature2020} to demonstrating spin-spin entanglement in a telecommunication network.\cite{knaut_entanglement_2023} 
In the context of quantum networks, color centers have also been proposed as transducers for scaling quantum computers based on, for example, superconducting qubits.\cite{kim_diamond_2023}   

Color centers are of course not only a potential building block to scale quantum hardware based on already established systems, but they are a viable resource for quantum applications in their own right. The electronic spin provides optical access to a register of nuclear spins, which makes the electronic spin a quantum bus providing access to a small scale processing node based on nuclear \cite{BradleyPRX2019} or electronic spins.\cite{knowles_demonstration_2016} Most recently a spin-network of 50 nuclear spin has been mapped by the authors of \cite{van_de_stolpe_mapping_2023} by using spin-echo double resonance (SEDOR). Similarly SEDOR was used to map electronic spins \cite{ungar_identification_2023} in an NV's environment. Both experiments show that the building blocks for small to intermediate scale quantum processors based on color centers in diamond are actively being developed.

Color centers are also viable for generating resource states \cite{lee_quantum_2019, istrati_sequential_2020, thomas_efficient_2022, cogan_deterministic_2023} for second and third generation quantum repeater architectures, which make use of so called photonic graph states.\cite{buterakos_deterministic_2017, BorregaardPRX2019, wo_resource-efficient_2023} In the same vein they can be used to generate the discrete variable cluster states, that have been proposed as a resource for measurement based fault tolerant quantum computing.\cite{nielsen_fault-tolerant_2005,briegel_measurement-based_2009}

Long-term coherence maintenance in quantum memories or registers is another goal, enabling permanent quantum state storage. This advancement will boost quantum memories' applications in quantum technologies, such as secure quantum tokens \cite{Pastawski2012} and quantum money,\cite{bozzio_experimental_2018} which are more reliant on long coherence times. Recent demonstrations of minute-long storage times \cite{BradleyPRX2019} suggest that permanent quantum state storage is becoming more feasible.

In conclusion, despite ongoing challenges, diamond color centers hold great promise for quantum applications. This review has outlined various quantum control approaches and examined their state-of-the-art performance. Coherent manipulation of spin qubits in diamond enabled entanglement generation and fundamental quantum network applications. Advancing these applications requires integrated photonics to allow for device scaling while, at the same time, preserving the optical and spin coherence properties. A major challenge is the charge and spin noise affecting shallow color centers near surfaces, causing spectral diffusion and spin dephasing. Another challenge is experimental complexity; for instance, the well-established SiV center requires operation at millikelvin temperatures in dilution refrigerators. Heavier color centers like SnV and PbV could operate at cryogenic temperatures, but their ability to generate spin-photon entanglement is not yet proven. In summary, diamond color centers offer potential for both local and remote quantum operations, leveraging both spin and photonic states. With continued rapid progress, the implementation of large-scale quantum networks and quantum information processing using color centers is a realistic goal.

\medskip
\textbf{Acknowledgements} \par 
The authors acknowledge funding by the European Research Council (ERC, Starting Grant project QUREP, No. 851810), the German Federal Ministry of Education and Research (BMBF, project DiNOQuant, No. 13N14921; project QPIS, No. 16KISQ032K; project QPIC-1, No. 13N15858; project QR.X, No. KIS6QK4001), and the Einstein Foundation Berlin (Einstein Research Unit on Quantum Devices).

\medskip

\bibliography{review_arxiv.bib}

\end{document}